\documentclass[12pt,english]{article}
\usepackage[T1]{fontenc}
\usepackage[latin9]{inputenc}
\usepackage{geometry}
\usepackage{enumitem}
\usepackage{amsmath}
\usepackage{amsthm}
\usepackage{amssymb}
\usepackage{stackrel}
\usepackage{setspace}
\makeatletter
\IfFileExists{candleshape.def}{%
 \input{candleshape.def}}{}
\IfFileExists{dropshape.def}{%
 \input{dropshape.def}}{}
\IfFileExists{TeXshape.def}{%
 \input{TeXshape.def}}{}
\IfFileExists{triangleshapes.def}{%
 \input{triangleshapes.def}}{}

\@ifundefined{lettrine}{\usepackage{lettrine}}{}
\numberwithin{equation}{section}
\numberwithin{figure}{section}
\numberwithin{table}{section}
\theoremstyle{definition}
\newtheorem{defn}{\protect\definitionname}
\theoremstyle{remark}
\newtheorem{rem}{\protect\remarkname}
\theoremstyle{plain}
\newtheorem{thm}{\protect\theoremname}
\theoremstyle{definition}
\newtheorem{example}{\protect\examplename}
\theoremstyle{plain}
\newtheorem{prop}{\protect\propositionname}

\makeatother

\usepackage{babel}
\providecommand{\definitionname}{Definition}
\providecommand{\examplename}{Example}
\providecommand{\propositionname}{Proposition}
\providecommand{\remarkname}{Remark}
\providecommand{\theoremname}{Theorem}

\begin{document}
\title{Dynamic Reserves in Matching Markets\thanks{First version: September, 2016. This version: March, 2020. \protect \\
We are grateful to the editor, the associate editor, and two anonymous
referees, as well as Péter Biró, Fuhito Kojima, Scott Duke Kominers,
Utku Ünver, Rakesh Vohra, and especially Bumin Yenmez, whose detailed
suggestions lead to significant improvements in the paper. We also
thank Dilip Abreu, O\u{g}uz Afacan, Nikhil Agarwal, Eduardo Azevedo,
Jenna M. Blochowicz, Rahul Deb, Federico Echenique, Isa Hafal\i r,
Andrei Gomberg, John W. Hatfield, Yuichiro Kamada, Navin Kartik, Onur
Kesten, Parag Pathak, Ran Shorrer, Tayfun Sönmez, and Alexander Westkamp
for helpful comments. Finally, we thank the seminar audience at Boston
College, University of St. Andrews, ITAM, 2017 MATCH-UP Conference,
2017 AEA Meeting, 2016 North American Meeting of Econometric Society,
2016 GAMES,  the 13th Meeting of Social Choice and Welfare, 2015 Conference
on Economic Design, 9th Meeting of Matching in Practice, and Workshop
on Market Design in memory of Dila Meryem Hafal\i r at ITAM. }}
\author{Orhan Aygün\thanks{orhan.aygun@boun.edu.tr; Bo\u{g}aziçi University, Department of Economics,
Natuk Birkan Building, Bebek, Istanbul 34342, Turkey. } $\quad$and $\quad$Bertan Turhan\thanks{bertan@iastate.edu; Iowa State University, Department of Economics,
Heady Hall, 518 Farm House Lane, Ames, IA 50011, USA.}}
\date{March, 2020\bigskip{}
}
\maketitle
\begin{abstract}
We study a school choice problem under affirmative action policies
where authorities reserve a certain fraction of the slots at each
school for specific student groups, and where students have preferences
not only over the schools they are matched to but also the type of
slots they receive. Such reservation policies might cause waste in
instances of low demand from some student groups. To propose a solution
to this issue, we construct a family of choice functions, \emph{dynamic
reserves choice functions, }for schools that respect within-group
fairness and allow the transfer of otherwise vacant slots from low-demand
groups to high-demand groups. We propose the cumulative offer mechanism
(COM) as an allocation rule where each school uses a dynamic reserves
choice function and show that it is \emph{stable} with respect to
schools' choice functions, is \emph{strategy-proof}, and\emph{ respects
improvements}. Furthermore, we show that transferring more of the
otherwise vacant slots leads to strategy-proof Pareto improvement
under the COM. 

\bigskip{}

\emph{$\mathbf{JEL}$ $\mathbf{Classification}$: }C78, D47, D61,
D63\vfill{}
\end{abstract}
\pagebreak{}

\section{Introduction}

The theory of two-sided matching and its applications has been studied
since the seminal work of \emph{Gale and Shapley} (1962). Nevertheless,
many real-life matching markets are subject to various constraints,
such as affirmative action in school choice. Economists and policy
makers are often faced with new challenges from such constraints.
Admission policies in school choice systems often use reserves to
grant applicants from certain backgrounds higher priority for some
available slots. \emph{Reservation in India }is such a process of
setting aside a certain percentage of slots in government institutions
for members of underrepresented communities, defined primarily by
castes and tribes. We present engineering school admissions in India
as an unprecedented matching problem with affirmative action in which
students care about the category through which they are admitted.

In engineering school admissions in India, students from different
backgrounds (namely, \emph{scheduled castes} (SC), \emph{scheduled
tribes} (ST), \emph{other backward classes} (OBC), and \emph{general
category} (GC))\footnote{Students who do not belong to SC, ST, and OBC categories are called
general category (GC) applicants.} are treated with different criteria. Schools reserve a certain fraction
of their slots for students from SC, ST, and OBC categories. The remaining
slots at each school, which are called \emph{general category} (GC)
slots, are open to competition. It is \emph{optional }for\emph{ }SC,
ST, and OBC students to declare their background information. Those
who do declare their background information are considered for the
reserved slots in their respective category, as well as for the GC
slots. Students who do not belong to SC, ST, or OBC categories are
considered only for GC slots. Students belonging to SC, ST, and OBC
communities who do not reveal their background information are only
considered for GC slots. \emph{Aygün and Turhan }(2017) documented
that students from SC, ST, and OBC categories have preferences not
only for schools but also for the category through which they are
admitted. Hence, students from these communities may prefer not to
declare their caste and tribe information in the application process.
Besides this strategic calculation burden on students, the current
admission procedure\footnote{Admission to the Indian Institute of Technologies (IITs) and its matching-theoretical
shortcomings are explained in detail in \emph{Aygün and Turhan} (2017). } suffers from a crucial market failure: The assignment procedure fails
to transfer some unfilled slots reserved for under-privileged castes
and tribes to the use of remaining students. Hence, it is quite wasteful. 

We address real-life applications as follows: There are schools and
students to be matched. Each school initially reserves a certain number
of its slots for different privilege groups (or student types). A
given student may possibly match with a given school under more than
one type. Each school has a pre-specified sequence\footnote{We will call this sequence a \emph{precedence sequence, }which is
different than the \emph{precedence order }from \emph{Kominers and
Sönmez} (2016). Precedence order is a linear order over the set of
student types. Precedence sequence, on the other hand, is more general
in the sense that a given student type might appear multiple times.
A technical definition will be given in the model section. } in which different sets of slots are considered, and where each set
accepts students in a single privilege type. Different schools might
have different orders. Since a student might have more than one privilege
type, the set of students cannot be partitioned into privilege groups.
 Each student has a preference over school-privilege type pairs. Students
care not only about which school they are matched to but also about
the privilege type under which they are admitted. Each school has
a target distribution of its slots over privilege types, but they
do not consider these target distributions as hard bounds\footnote{Hard bounds and soft bounds are analyzed in detail in \emph{Hafal\i r
et al. }(2013) and \emph{Ehlers et al.} (2014).}. If there is less demand from at least one privilege type, schools
are given the opportunity to utilize vacant slots by transferring
them over to other privilege types. Authorities might require a certain
capacity transfer scheme so that each school has a complete plan where
they state how they want to redistribute these slots. Thus, we take
capacity transfer schemes exogenously given. The only mild condition
imposed on the capacity transfer scheme is \emph{monotonicity,}\footnote{\emph{Westkamp} (2013) introduces this monotonicity condition on capacity
transfer schemes.}\emph{ }which requires that (1) if more slots are left from one or
more sets, the capacity of the sets considered later in the precedence
sequence must be weakly higher, and (2) a school cannot decrease the
total capacity in response to increased demand for some sets of slots. 

We design choice functions for schools that allow them to transfer
capacities from low-demand privilege types to high-demand privilege
types. Each school respects an exogenously given precedence sequence
between different sets of slots when it fills its slots. Each school
has a strict priority ordering (possibly different than the other
schools') over all students. For each school, priority orderings for
different privilege types are straightforwardly derived from the school's
priority ordering. There is an associated choice function, which we
call a \emph{``sub-choice function},'' for each set of slots. In
Indian engineering school admissions, sub-choice functions are \emph{q-responsive.}
That is, a sub-choice function always selects the q-best students
with respect to the priority ordering of the associated privilege
type at that school, where q denotes the capacity. 

The school starts filling its first set of slots according to its
precedence sequence. Given the initial capacity of the first set of
slots and a contract set, the sub-choice function associated with
the first set selects contracts. The school then moves to the second
set according to its precedence sequence. The (dynamic) capacity of
the second set is a function of the number of unfilled slots in the
first set. The exogenous capacity transfer function of the school
specifies the capacity of the second set. The set of available contracts
for the second set of slots is computed as follows: If a student has
one of her contracts chosen by the first set, then all of her contracts
are removed for the rest of the choice process. Given the set of remaining
contracts and the capacity, the sub-choice function associated with
the second set selects contracts. In general, the (dynamic) capacity
of set $k$ is a function of the number of vacant slots of the $k-1$
sets that precede it. The set of contracts available to the set of
slots $k$ is computed as follows: If a student has one of her contracts
chosen by one of the $k-1$ sets of slots that precede the $k^{th}$
set, then all of her contracts are removed. Given the set of remaining
contracts for the set of slots $k$ and its capacity, the sub-choice
function associated with the set $k$ selects contracts. The (overall)
choice of a school is the union of sub-choices of its sets of slots. 

We propose a remedy for the Indian engineering school admissions problem
through a matching with contracts model that has the ability to utilize
vacant slots of certain types for other students. We have three design
objectives: \emph{stability, strategy-proofness }and \emph{respect
for improvements. }Stability ensures that $(1)$ no student is matched
with an unacceptable school-slot category pair, $(2)$ schools' dynamic
reserves choices are respected, and $(3)$ no student desires a slot
at which she has a justified claim under the priority and precedence
structure. Strategy-proofness guarantees that students can never game
the allocation mechanism via preference manipulation. In our framework,
it also relieves students of the strategic manipulation burden, which
involves whether or not students declare their background.\footnote{Strategy-proofness ensures that it is a weakly dominant strategy for
each student to report their caste and tribe information.} Respect for improvements\footnote{See \emph{Kominers }(2019) for detailed discussion of respect for
improvements in matching markets.} is an essential property in meritocratic systems. In allocation mechanisms
that respect improvements, students have no incentive to lower their
standings in schools' priority rankings. 

We propose the cumulative offer mechanism (COM) as an allocation rule.
We prove that the COM is stable with respect to schools' dynamic reserve
choice functions\emph{ }(Theorem 1)\emph{, }is (weakly) group strategy-proof\emph{
}(Theorem 2)\emph{, }and respects improvements (Theorem 3). The main
result of the paper\emph{ }(Theorem 4) states that when a single school's
choice function becomes \emph{``more flexible,''}\footnote{We define \emph{``more flexible''} criterion to compare two monotonic
capacity transfer schemes given a precedence sequence. We say that
a monotonic capacity transfer scheme $\widetilde{q}$ is more flexible
than monotonic capacity transfer scheme $q$ if $\widetilde{q}$ transfers
at least as many otherwise vacant slots as $q$ at every instance.
There must also be an instance where $\widetilde{q}$ transfers strictly
more otherwise vacant slots than $q$ does.} while those of the other schools remain unchanged, the outcome of
the COM under the former (weakly) Pareto dominates the outcome under
the latter. Theorem 4 is of particular importance because it describes
a strategy-proof Pareto improvement. Finally, we investigate the relationship
between families of dynamic reserves choice rules and Kominers and
Sönmez's (2016) slot-specific priorities choice rules. We show that
for every slot-specific priorities choice rule, there is an outcome
equivalent dynamic reserves choice rule (Theorem 5). Moreover, we
give an example of a dynamic reserves choice rule for which there
is no outcome equivalent slot-specific priorities choice rule (Example
1). 

\subsection*{Related Literature}

The school choice problem was first introduced by the seminal paper
of \emph{Abdulkadiro\u{g}lu and Sönmez} (2003). The authors introduced
a simple affirmative action policy with\emph{ }type-specific quotas.
\emph{Kojima} (2012) showed that the minority students who purported
to be the beneficiaries might instead be made worse off under this
type of affirmative action. To circumvent inefficiencies caused by
majority quotas, \emph{Hafal\i r et al.} (2013) offer minority reserves\emph{.
Westkamp }(2013) introduced a model of matching with complex constraints.
His model permits priorities to vary across slots. In his model, students
are considered to be indifferent between different slots of a given
school. However, in our framework, students have strict preferences
for type-specific matches with schools. This crucial aspect differentiates
our paper from \emph{Westkamp} (2013). Moreover, our comparative statics
result on transfer schemes does not have a counterpart in \emph{Westkamp}
(2013). 

\emph{Kominers and Sönmez} (2016) introduce another prominent family
of choice functions---slot-specific priorities choice functions---to
implement diversity objectives in many-to-one settings. We show that
dynamic reserves choice rules nest slot-specific priorities choice
rules. Moreover, we provide an example of a dynamic reserves choice
rule that cannot be generated by a slot-specific priorities choice
rule. 

In a related work, \emph{Biró et al.} (2010) analyze a college admission
model with common and upper quotas in the context of Hungarian college
admissions. They use choice functions for colleges that allow them
to select multiple contracts of the same applicant. They show that
a stable assignment exists. The completions of dynamic reserves choice
functions, discussed in Appendix 7.2, satisfy the properties they
impose. Hence, their result also implies the existence of a stable
allocation in our framework. However, our main focus is different
as we aim to show strategy-proof Pareto improvement by making capacity
transfer function more flexible. 

The matching problem with dynamic reserves choice functions is a special
case of the matching with contracts\emph{ }model of \emph{Fleiner}
(2003)\footnote{Fleiner's results cover these of \emph{Hatfield and Milgrom} (2005)
regarding stability. However,\emph{ Fleiner} (2003) does not analyze
incentives.} and \emph{Hatfield and Milgrom} (2005).\footnote{\emph{Echenique} (2012) has shown that under the substitutes condition,
which is thoroughly assumed in \emph{Hatfield and Milgrom} (2005),
the matching with contracts model can be embedded within the \emph{Kelso
and Crawford} (1982) labor market model. \emph{Kelso and Crawford}
(1982) built on the analysis of \emph{Crawford and Knoer} (1981).} The analysis and results of \emph{Hatfield and Kominers} (2019) are
the technical backbone of our results regarding stable and strategy-proof
mechanism design. We show that every dynamic reserves choice function
has a completion that satisfies the irrelevance of rejected contracts
condition of \emph{Aygün and Sönmez} (2013), in conjunction with substitutability
and the law of aggregate demand. 

\emph{Hatfield et al. }(2017) introduce a model of hospital choice
in which each hospital has a set of divisions and \emph{flexible allotment
of capacities }to those divisions that vary as a function of the set
of contracts available. These authors define choice functions that
nest dynamic reserves choice functions while continuing to obtain
stability and strategy-proofness for the COM. Our Theorems 3 and 4
do not have a counterpart in \emph{Hatfield et al. }(2017). 

Our work is also related with the research agenda on matching with
constraints that is studied in a series of papers: \emph{Kamada and
Kojima} (2015), (2017), \emph{Kojima et al.} (2018), and \emph{Goto
et al.} (2017). In these papers, constraints are imposed on subsets
of institutions as a joint restriction, as opposed to at each individual
institution. Our main results distinguish our work from these papers.
We discuss the relationship between our stability notion and that
of \emph{Kamada and Kojima }(2017) in Section 3. 

Another related paper is \emph{Echenique and Yenmez} (2015). Dynamic
reserves choice functions might seem similar to the family of choice
functions the authors analyze: choice rules generated by reserves\emph{.
}However, dynamic reserves choice functions choose contracts whereas
choice rules generated by reserves choose students. 

Two recent papers, \emph{Sönmez and Yenmez }(2019a,b)\emph{, }study
affirmative action in India from a matching-theoretical perspective.
The authors consider both vertical and horizontal reservations\footnote{Caste-based reservations for SC, ST, and OBC categories are called
vertical reservations, also referred to as social reservations. Horizontal
reservations, also referred to as special reservations, are intended
for other disadvantaged groups of citizens, such as disabled persons,
and women. Horizontal reservations are implemented within each vertical
category. See \emph{Sönmez and Yenmez }(2019a,b) for details.} while we consider only vertical reservations for simplicity. Even
though they consider more general reserve structure than ours, the
authors consider agents' preferences only over institutions and do
not take agents' preferences over vertical categories they are admitted
under into account. Moreover, they assume away capacity transfers
between vertical categories. Therefore, their model does not contain
our model and vice versa. 

\section{Model}

There is a finite set of students $I=\{i_{1},...,i_{n}\}$, a finite
set of schools $S=\{s_{1},...,s_{m}\}$, and a finite set of student
privileges (types)\footnote{We use the terms \emph{``type'' }and \emph{``privilege'' }interchangeably.}
$T=\{t_{1},...,t_{p}\}$. We call $T_{i}\subseteq T$ the set of privileges
that student $i$ can claim and $\mathbf{T}=(T_{i})_{i\in I}$ the
profile of types that students can claim. We define $X_{i}=\{i\}\times S\times T_{i}$
as the set of all contracts associated with student $i\in I$. We
let $X=\underset{i\in I}{\cup}X_{i}$ be the set of all contracts.
Each contract $x\in X$ is between a student $\mathbf{i}(x)$ and
a school $\mathbf{s}(x)$ and specifies a privilege $\mathbf{t}(x)\in T_{\mathbf{i}(x)}$.
There may be many contracts for each student-school pair. We extend
the notations $\mathbf{i}(\cdot)$, $\mathbf{s}(\cdot)$ and $\mathbf{t}(\cdot)$
to the set of contracts for any $Y\subseteq X$ by setting $\mathbf{i}(Y)\equiv\underset{y\in Y}{\cup}\{\mathbf{i}(y)\}$,
$\mathbf{s}(Y)\equiv\underset{y\in Y}{\cup}\{\mathbf{s}(y)\}$ and
$\mathbf{t}(Y)\equiv\underset{y\in Y}{\cup}\{\mathbf{t}(y)\}$. For
$Y\subseteq X$, we denote $Y_{i}\equiv\{y\in Y\mid\mathbf{i}(y)=i\}$;
analogously, we denote $Y_{s}\equiv\{y\in Y\mid\mathbf{s}(y)=s\}$
and $Y_{t}\equiv\{y\in Y\mid\mathbf{t}(y)=t\}$. 

Each student $i\in I$ has a (linear) preference order $P^{i}$ over
contracts in $X_{i}=\{x\in X\mid\mathbf{i}(x)=i\}$ and an \emph{outside
option }$\emptyset$ which represents remaining unmatched. A contract
$x\in X_{i}$ is\emph{ acceptable} for $i$ (with respect to $P^{i}$)
if $xP^{i}\emptyset$. We use the convention that $\emptyset P^{i}x$
if $x\in X\setminus X_{i}$. We say that the contracts $x\in X$ for
which $\emptyset P^{i}x$ are \emph{unacceptable }to \emph{i . }The
\emph{at-least-as-well }relation $R^{i}$ is obtained from $P^{i}$
as follows: $xR^{i}x^{'}$ if and only if either $xP^{i}x^{'}$ or
$x=x^{'}$. Let $\mathcal{P}^{i}$ denote the set of all preferences
over $X_{i}\cup\{\emptyset\}$. A preference profile of students is
denoted by $P=(P^{i_{1}},...,P^{i_{n}})\in\times_{i\in I}\mathcal{P}^{i}$.
A preference profile of all students except student $i_{l}$ is denoted
by $P_{-i_{l}}=(P^{i_{1}},...,P^{i_{l-1}},P^{i_{l+1}},...,P^{i_{n}})\in\times_{i\neq i_{l}}\mathcal{P}^{i}$. 

Students have \emph{unit demand}, that is, they choose at most one
contract from a set of contract offers. We assume that students always
choose the best available contract, so that the choice $C^{i}(Y)$
of a student $i\in I$ from contract set $Y\subseteq X$ is the $P^{i}$-maximal
element of $Y$ (or the outside option if $\emptyset P^{i}y$ for
all $y\in Y$).\footnote{To simplify our notation, the individual contracts are treated as
interchangeable with singleton contract sets.}

For each school $s\in S$, $\overline{q}_{s}$ denotes the physical
capacity of school $s\in S$. We call $\overline{q}=(\overline{q}_{s_{1}}...,\overline{q}_{s_{m}})$
the vector of school capacities. Each school $s\in S$ has a priority
order $\pi^{s}$, which is a linear order over $I\cup\{\emptyset\}$.\footnote{This priority order is often determined by performance on an admission
exam, by a random lottery, or dictated by law. In engineering school
admissions in India, each school ranks students according to test
scores. Different schools might have different test score rankings
because they use different weighted averages of math, physics, chemistry,
and biology scores depending on the school. It is important to note
that students whose test scores are under a certain threshold are
deemed as unacceptable for each school. } Let $\Pi=(\pi^{s_{1}},...,\pi^{s_{m}})$ denote the priority profile
of schools. For each school $s\in S$, the priority ordering for students
who can claim the privilege $t\in T$, denoted by $\pi_{t}^{s}$,
is obtained from $\pi^{s}$ as follows:
\begin{itemize}
\item for $i,j\in I$ such that $t\in T_{i}\setminus T_{j}$, $i\pi^{s}\emptyset$,
and $j\pi^{s}\emptyset$,  $i\pi_{t}^{s}\emptyset\pi_{t}^{s}j$,\footnote{$\emptyset\pi_{t}^{s}j$ means student $j$ is unacceptable for privilege
$t$ at school $s$.}
\item for any other $i,j\in I$, $i\pi_{t}^{s}j$ if and only if $i\pi^{s}j$. 
\end{itemize}
An \emph{allocation }$Y\subseteq X$ is a set of contracts such that
each student appears in \emph{at most one }contract and no school
appears in more contracts than its capacity allows. Let $\mathcal{X}$
denote the set of all allocations. Given a student $i$ and an allocation
$Y$, we refer to the pair $(\mathbf{s}(x),\mathbf{t}(x))$ such that
$\mathbf{i}(x)=i$ as the \emph{assignment }of student $i$ under
allocation $Y$. We extend student preferences over contracts to preferences
over outcomes in the natural way. We say that an allocation $Y\subseteq X$
\emph{Pareto dominates }allocation $Z\subseteq X$ if $Y_{i}R^{i}Z_{i}$
for all $i\in I$ and $Y_{i}P^{i}Z_{i}$ for at least one $i\in I$. 

\subsection{Dynamic Reserves Choice Functions}

Each school $s\in S$ has multi-unit demand, and is endowed with a
choice function $C^{s}(\cdot)$ that describes how $s$ would choose
from any offered set of contracts. Throughout the paper, we assume
that for all $Y\subseteq X$ and for all $s\in S$, the choice function
$C^{s}(\cdot)$: 
\begin{enumerate}
\item only selects contracts to which $s$ is a party, i.e., $C^{s}(Y)\subseteq Y_{s}$,
and
\item selects at most one contract with any given student. 
\end{enumerate}
For any $Y\subseteq X$ and $s\in S$, we denote $R^{s}(Y)\equiv Y\setminus C^{s}(Y)$
as the set of contracts that $s$ \emph{rejects }from $Y$. 

We now introduce a model of dynamic reserves choice functions in which
each school $s\in S$ has $\lambda_{s}$ \emph{groups of slots}. School
$s$ fills its groups of slots according to a \emph{precedence sequence,}\footnote{We take precedence sequences to be exogenously given. However, \emph{Dur
et al.} (2018) show that precedence sequences might have significant
effects on distributional objectives in the context of Boston's school
choice system.}\emph{ }which is a surjective function $f^{s}:\{1,...,\lambda_{s}\}\longrightarrow T$.
The interpretation of $f^{s}$ is that school $s$ fills the first
group of slots with $f^{s}(1)$-type students, the second group of
slots with $f^{s}(2)$-type students, and so on. School $s\in S$
has a \emph{target distribution }of its slots across different types
$(\overline{q}_{s}^{t_{1}},...,\overline{q}_{s}^{t_{p}})$, which
means that it has $\overline{q}_{s}^{t_{1}}$ slots to be reserved
for privilege $t_{1}$, $\overline{q}_{s}^{t_{2}}$ slots to be reserved
for privilege $t_{2}$, and so on. To satisfy its target reserve structure,
school $s$ fills its slots according to the initially set capacities
for each group of slots\emph{ }$(\overline{q}_{s}^{1},\overline{q}_{s}^{2},...,\overline{q}_{s}^{\lambda_{s}})$
such that $\underset{j\in(f^{s})^{-1}(t)}{\sum}\overline{q}_{s}^{j}=\overline{q}_{s}^{t}$
for all $t\in T$. If the target distribution cannot be achieved because
too few students from one or more privileges apply, then school $s$
use an exogenously given capacity transfer scheme that specifies how
its capacity is to be redistributed. Technically, a capacity transfer
scheme is defined as follows:
\begin{defn}
Given a precedence sequence $f^{s}$ and a capacity of the first group
of slots $\overline{q}_{s}^{1}$, a $\mathbf{capacity}$ $\mathbf{transfer}$
$\mathbf{scheme}$ of school $s$ is a sequence of capacity functions
$q_{s}=(\overline{q}_{s}^{1},(q_{s}^{k})_{k=2}^{\lambda_{s}})$, where
$q_{s}^{k}:\:\mathbb{Z}_{+}^{k-1}\longrightarrow\mathbb{Z}_{+}$ such
that $q_{s}^{k}(0,...,0)=\overline{q}_{s}^{k}$ for all $k\in\{2,...,\lambda_{s}\}$. 
\end{defn}
We impose a mild condition, à la \emph{Westkamp} (2013), on capacity
transfer functions. 
\begin{defn}
A capacity transfer scheme $q_{s}$ is $\mathbf{monotonic}$ if, for
all $j\in\{2,...,\lambda_{s}\}$ and all pairs of sequences $(r_{l},\widetilde{r}_{l})$
such that $\widetilde{r}_{l}\geq r_{l}$ for all $l\leq j-1$, 
\end{defn}
\begin{itemize}
\item $q_{s}^{j}(\widetilde{r}_{1},...,\widetilde{r}_{j-1})\geq q_{s}^{j}(r_{1},...,r_{j-1})$,
and 
\item $\stackrel[m=2]{j}{\sum}[q_{s}^{m}(\tilde{r}_{1},...,\widetilde{r}_{m-1})-q_{s}^{m}(r_{1},...,r_{m-1})]\leq\stackrel[m=1]{j-1}{\sum}[\widetilde{r}_{m}-r_{m}]$. 
\end{itemize}
Monotonicity of capacity transfer schemes requires that (1) whenever
weakly more slots are left unfilled in \emph{every }groups of slots
preceding the $j^{th}$ group of slots, weakly more slots should be
available for the $j^{th}$ group, and (2) a school cannot decrease
its total capacity in response to increased demand for some groups
of slots. 

\subsection*{Sub-choice functions}

For each group of slots at school $s\in S$, there is an associated
\emph{sub-choice} function $c^{s}:\:2^{X}\times\mathbb{Z}_{+}\times T\longrightarrow2^{X}$.
Given a set of contracts $Y\subseteq X$, a nonnegative integer $\kappa\in\mathbb{Z}_{+}$,
and a privilege $t\in T$, $c^{s}(Y,\kappa,t)$ denotes the set of
chosen contracts that name privilege $t$ up to the capacity $\kappa$
from the set of contracts $Y$. We require sub-choice functions to
be \emph{q-responsive }given the ranking $\pi_{t}^{s}$.
\begin{defn}
\footnote{We adapt this definition from \emph{Chambers and Yenmez} (2017).}A
sub-choice function $c^{s}(\cdot,\kappa,t)$ of a group of slots at
school $s$ for privilege type $t$ is \emph{q-responsive }if there
exists a strict priority ordering $\pi_{t}^{s}$ on the set of contracts
naming privilege type $t$ and a positive integer $\kappa$, such
that for any $Y\subseteq(X_{s}\cap X_{t})$, 

\[
c^{s}(Y,\kappa,t)=\stackrel[i=1]{\kappa}{\cup}\{y_{i}^{*}\}
\]
 where $y_{i}^{*}$ is defined as $y_{1}^{*}=\underset{\pi_{t}^{s}}{max}Y$
and, for $2\leq i\leq\kappa$, $y_{i}^{*}=\underset{\pi_{t}^{s}}{max}Y\setminus\{y_{1}^{*},...,y_{i-1}^{*}\}$
. 
\end{defn}
In other words, a sub-choice function is q-responsive\footnote{These types of sub-choice functions are often used in real-life applications.
For example, in the cadet branch matching processes in the USMA and
ROTC, each sub-choice function is induced from a strict ranking of
students according to test scores. See \emph{Sönmez and Switzer }(2013)
and \emph{Sönmez }(2013) for further details.} if there is a strict priority ordering over students who have privilege
$t$ for which the sub-choice function always selects the highest-ranked
available students in privilege $t$ up to the capacity. 
\begin{rem}
Since our main real-life application is engineering school admissions
in India, we shall assume that at each school $s\in S$, and for each
group of slots reserved for privilege $t\in T$, the associated sub-choice
function $c^{s}(\cdot,\cdot,t)$ is q-responsive and obtained from
$\pi_{t}^{s}$. 
\end{rem}

\subsection*{Overall choice functions}

The $\mathbf{overall}$ $\mathbf{choice}$ $\mathbf{function}$ of
school $s$, $C^{s}(\cdot,f^{s},q_{s}):\:2^{X}\longrightarrow2^{X}$,
runs its sub-choice functions in an orderly fashion given the precedence
sequence $f^{s}$ and capacity transfer scheme $q_{s}$. Given a set
of contracts $Y\subseteq X$, $C^{s}(Y,f^{s},q_{s})$ denotes the
set of chosen contracts from the set of contracts $Y$ and is determined
as follows: 
\begin{itemize}
\item Given $\overline{q}_{s}^{1}$ and $Y=Y^{0}\subseteq X$, let $Y_{1}\equiv c_{1}^{s}(Y^{0},\overline{q}_{s}^{1},f^{s}(1))$
be the set of chosen contracts with privilege $f^{s}(1)$ from $Y^{0}$.
Let $r_{1}=\overline{q}_{s}^{1}-\mid Y_{1}\mid$ be the number of
vacant slots. Define $\widetilde{Y}_{1}\equiv\{y\in Y^{0}\mid\mathbf{i}(y)\in\mathbf{i}(Y_{1})\}$
as the set of all contracts of students whose contracts are chosen
by sub-choice function $c_{1}^{s}(\cdot,\overline{q}_{s}^{1},f^{s}(1))$.
If a contract of a student is chosen, then all of the contracts naming
that student shall be removed from the set of available contracts
for the rest of the procedure. The set of remaining contracts is then
$Y^{1}=Y^{0}\setminus\widetilde{Y}_{1}$. 
\item In general, let $Y_{k}=c_{k}^{s}(Y^{k-1},q_{s}^{k},f^{s}(k))$ be
the set of chosen contracts with privilege $f^{s}(k)$ from the set
of available contracts $Y^{k-1}$ , where $q_{s}^{k}=q_{s}^{k}(r_{1},...,r_{k-1})$
is the dynamic capacity of group of slots $k$ as a function of the
vector of the number of unfilled slots $(r_{1},...,r_{k-1})$. Let
$r_{k}=q_{s}^{k}-\mid Y_{k}\mid$ be the number of vacant slots. Define
$\widetilde{Y}_{k}=\{y\in Y^{k-1}\mid\mathbf{i}(y)\in\mathbf{i}(Y_{k})\}$.
The set of remaining contracts is then $Y^{k}=Y^{k-1}\setminus\widetilde{Y}_{k}$. 
\item Given $Y=Y^{0}\subseteq X$ and the capacity of the first group of
slots $\overline{q}_{s}^{1}$ , we define the \emph{overall choice
function }of school $s$ as $C^{s}(Y,f^{s},q_{s})=c_{1}^{s}(Y^{0},\overline{q}_{s}^{1},f^{s}(1))\cup(\stackrel[k=2]{\lambda_{s}}{\cup}c_{k}^{s}(Y^{k-1},q_{s}^{k}(r_{1},...,r_{k-1}),f^{s}(k)))$. 
\end{itemize}
The primitives of the overall choices for each school $s\in S$ are
the precedence sequence $f^{s}$, the capacity transfer scheme $q_{s}$,
and the priority order $\pi^{s}$. Since an overall choice is computed
by using these primitives, it is not one of the primitives in our
model. The list $\left(I,S,\mathbf{T},X,P,\Pi,(f^{s},q_{s},\pi^{s})_{s\in S}\right)$
denotes a problem. 

\section{Stability Concept}

Stability has emerged as the key to a successful matching market design.
We follow the\emph{ Gale and Shapley} (1962) tradition in focusing
on outcomes that are\emph{ }stable\emph{. }In the matching with contracts
framework, \emph{Hatfield and Milgrom} (2005) define stability as
follows: An outcome $Y\subseteq X$ is $\mathbf{stable}$ if 
\begin{enumerate}
\item $Y_{i}R^{i}\emptyset$ for all $i\in I$, 
\item $C^{s}(Y)=Y_{s}$ for all $s\in S$, and
\item there does not exist a school $s\in S$ and a $\mathbf{blocking}$
$\mathbf{set}$ $Z\neq C^{s}(Y)$ such that $Z_{s}\subseteq C^{s}(Y\cup Z)$
and $Z_{i}=C^{i}(Y\cup Z)$ for all i$\in\mathbf{i}(Z)$. 
\end{enumerate}
If the first requirement (\emph{individual rationality for students})
fails, then there is a student who prefers to reject a contract that
involves her (or, equivalently, there is a student who is given an
unacceptable contract). In our context, the second condition (\emph{individual
rationality for schools}) requires that the schools' choice functions
are respected. If the third condition\emph{ }(\emph{unblockedness})
fails, then there is an alternative set of contracts that a school
and students associated with a contract in that set strictly prefers. 
\begin{rem}
Our stability notion is related to the \emph{weak stability }notion
of\emph{ Kamada and Kojima} (2017). The authors define the \emph{feasibility
constraint} as a map $\phi:\mathbb{Z}_{+}^{\mid H\mid}\longrightarrow\{0,1\}$,
such that $\phi(w)\geq\phi(w^{'})$ whenever $w\leq w^{'}$. Their
interpretation is that each coordinate in $w$ corresponds to a hospital
and the number in that coordinate represents the number of doctors
matched to that hospital. $\phi(w)=1$ means that $w$ is feasible
and $\phi(w)=0$ means it is not. They say that matching $\mu$ is
\emph{feasible }if and only if $\phi(w(\mu))=1$, where $w(\mu):=(\mid\mu_{h}\mid)_{h\in H}$
is a vector of nonnegative integers indexed by hospitals whose coordinates
corresponding to $h$ are $\mid\mu_{h}\mid$. Capacity transfer functions
in our setting can be represented by\emph{ the} feasibility constraint
map from their paper. Condition 2 in our stability definition takes
into account not only dynamic capacities of groups of seats in each
school but also their precedence sequences. It is a feasibility condition.
\emph{Westkamp} (2013) defines a similar condition in his \emph{``procedural
stability'' }definition in a simpler matching model without contracts. 
\end{rem}

\section{The Cumulative Offer Mechanism and its Properties under Dynamic Reserves
Choice Functions}

A \emph{direct mechanism }is a mechanism where the strategy space
is the set of preferences $\mathcal{P}$ for each student $i\in I$,
i.e., a function $\psi:\mathcal{P}^{n}\longrightarrow\mathcal{X}$
that selects an allocation for each preference profile. We propose
the COM  as our allocation function. Given the student preferences
and schools' overall choice functions, the outcome of the COM is computed
by the \emph{cumulative offer algorithm}. This is the generalization
of the agent-proposing deferred acceptance algorithm of \emph{Gale
and Shapley} (1962). We now introduce the cumulative offer process
(COP)\footnote{See \emph{Hatfield and Milgrom} (2005) for more details.}
for matching with contracts. Here, we provide an intuitive description
of this algorithm; we give a more technical description in Appendix
7.1. 
\begin{defn}
In the COP,  students propose contracts to schools in a sequence of
steps $l=1,2\ldots$ :

\emph{Step 1 }: Some student $i^{1}\in I$ proposes his most-preferred
contract, $x^{1}\in X_{i^{1}}$. School $s\left(x^{1}\right)$ holds
$x^{1}$ if $x^{1}\in C^{s\left(x^{1}\right)}\left(\left\{ x^{1}\right\} \right)$,
and rejects $x^{1}$ otherwise. Set $A_{s\left(x^{1}\right)}^{2}=\left\{ x^{1}\right\} $,
and set $A_{s'}^{2}=\emptyset$ for each $s'\neq s\left(x^{1}\right)$;
these are the sets of contracts available to schools at the beginning
of Step 2. 

\emph{Step 2 }: Some student $i^{2}\in I$, for whom no school currently
holds a contract, proposes his most-preferred contact that has not
yet been rejected, $x^{2}\in X_{i^{2}}$. School $s\left(x^{2}\right)$
holds the contract in $C^{s\left(x^{2}\right)}\left(A_{s\left(x^{2}\right)}^{2}\cup\left\{ x^{2}\right\} \right)$
and rejects all other contracts in $A_{s\left(x^{2}\right)}^{2}\cup\left\{ x^{2}\right\} $;
schools $s'\neq s\left(x^{2}\right)$ continue to hold all contracts
they held at the end of Step 1. Set $A_{s\left(x^{2}\right)}^{3}=A_{s\left(x^{2}\right)}^{2}\cup\left\{ x^{2}\right\} $,
and set $A_{s'}^{3}=A_{s'}^{2}$ for each $s'\neq s\left(x^{2}\right)$. 

\emph{Step l }: Some student $i^{l}\in I$, for whom no school currently
holds a contract, proposes his most-preferred contact that has not
yet been rejected, $x^{l}\in X_{i^{l}}$. School $s\left(x^{l}\right)$
holds the contract in $C^{s\left(x^{l}\right)}\left(A_{s\left(x^{l}\right)}^{l}\cup\left\{ x^{l}\right\} \right)$
and rejects all other contracts in $A_{s\left(x^{l}\right)}^{l}\cup\left\{ x^{l}\right\} $;
schools $s'\neq s\left(x^{l}\right)$ continue to hold all contracts
they held at the end of Step $l-$1. Set $A_{s\left(x^{l}\right)}^{l+1}=A_{s\left(x^{l}\right)}^{l}\cup\left\{ x^{l}\right\} $,
and set $A_{s'}^{l+1}=A_{s'}^{l}$ for each $s'\neq s\left(x^{l}\right)$. 
\end{defn}
If at any time no student is able to propose a new contract---that
is, if all students for whom no contracts are on hold have proposed
all contract they find acceptable---then the algorithm terminates.
The outcome of the COP\emph{ }is the set of contracts held by schools
at the end of the last step before termination. 

In the COP, students propose contracts sequentially. Schools accumulate
offers, choosing at each step (according to their choice functions)
a set of contracts to hold from the set of all previous offers. The
process terminates when no student wishes to propose a contract. 

Given a preference profile of students $P=\left(P_{i}\right)_{i\in I}$
and a profile of choice functions for schools $C=\left(C^{s}\right)_{s\in S}$,
let $\Phi\left(P,C\right)$ denote the outcome of the COM. Let $\Phi_{i}\left(P,C\right)$
denote the assignment of student $i\in I$ and $\Phi_{s}\left(P,C\right)$
denote the assignment of school $s\in S$.
\begin{rem}
We do not explicitly specify the order in which students make proposals.
\emph{Hirata and Kasuya} (2014) show that in the matching with contracts
model, the outcome of the COP is \emph{order-independent }if the overall
choice function of every school satisfies the bilateral substitutability
(BLS) and the irrelevance of rejected contracts (IRC) conditions.
Dynamic reserves choice functions satisfy BLS and IRC. Hence, the
order-independence of the COP holds.
\end{rem}
A mechanism $\varphi$ is \emph{stable }if for every preference profile
$P\in\mathcal{P}^{\mid I\mid}$ the outcome $\varphi\left(P\right)$
is stable with respect to the schools' overall choice functions. Since
the COP  gives a stable outcome for every input if each school's capacity
transfer scheme is monotonic, the COM is a stable mechanism. 
\begin{thm}
The cumulative offer mechanism is stable with respect to dynamic reserves
choice functions. 
\end{thm}

\paragraph{Proof.}

See Appendix 7.3. 

To analyze the incentive properties of the COM when schools use dynamic
reserves choice functions, we first define standard strategy-proofness
and (weak) group strategy-proofness in relation to a direct mechanism.
\begin{defn}
A direct mechanism $\varphi$ is said to be \textbf{strategy-proof
}if there does not exist a preference profile $P$, a student $i\in I$,
and preferences $P_{i}'$ of student $i$ such that 
\[
\varphi_{i}\left(P_{i}',P_{-i}\right)P_{i}\varphi_{i}\left(P\right).
\]
\end{defn}
That is, no matter which student we consider, no matter what her true
preferences $P_{i}$ are, no matter what other preferences $P_{-i}$
other students report (true or not), and no matter which potential
``misrepresentation'' $P_{i}'$ student $i$ considers, a truthful
preference revelation is in her best interest. Hence, students can
never benefit from gaming the mechanism $\varphi$.
\begin{defn}
A direct mechanism $\varphi$ is said to be $\mathbf{weakly}$ \textbf{group
strategy-proof }if there is no preference profile $P$, a subset of
students $I'\subseteq I$, and a preference profile $\left(P_{i}\right)_{i\in I'}$
of students in $I'$ such that 
\[
\varphi_{i}\left(\left(P_{i}'\right)_{i\in I'},\left(P_{j}\right)_{j\in I\setminus I'}\right)P_{i}\varphi_{i}\left(P\right)
\]
for all $i\in I'$.
\end{defn}
That is, no subset of students can jointly misreport their preferences
to receive a strictly preferred outcome for every member of the coalition. 

\emph{Hatfield and Kominers} (2019) show that if schools' choice functions
have substitutable completions so that these completions satisfy the
LAD, then the COP  becomes weakly group strategy-proof.  
\begin{thm}
Suppose that each school uses a dynamic reserves choice function.
 Then, the cumulative offer mechanism is weakly group strategy-proof.
\end{thm}

\paragraph{Proof.}

See Appendix 7.3.

\subsection*{Respect for Unambiguous Improvements}

We say that priority profile $\overline{\Pi}$ is an \emph{unambiguous
improvement over priority profile $\Pi$ }for student $i\in I$\emph{
}if, for all schools $s\in S$, the following conditions hold:
\begin{enumerate}
\item For all $x\in X_{i}$ and $y\in\left(X_{I\setminus\{i\}}\cup\{\emptyset\}\right)$,
if $x\pi^{s}y$ then $x\overline{\pi}^{s}y$.
\item For all $y,z\in X_{I\setminus\{i\}}$, $y\pi^{s}z$ if and only if
$y\overline{\pi}^{s}z$. 
\end{enumerate}
That is, $\overline{\Pi}$ is an unambiguous improvement over priority
profile $\Pi$ for student $i$ if $\overline{\Pi}$ is obtained from
$\Pi$ by increasing the priority of some of $i$'s contracts while
leaving the relative priority of other students' contracts unchanged. 
\begin{defn}
A mechanism $\varphi$ \textbf{$\mathbf{respects}$ $\mathbf{unambiguous}$
$\mathbf{improvements}$} for $i\in I$ if for any preference profile
$P\in\times_{i\in I}\mathcal{P}^{i}$ 
\[
\varphi_{i}(P;\overline{\Pi})R^{i}\varphi_{i}(P;\Pi)
\]
 whenever $\overline{\Pi}$ is an unambiguous improvement over $\Pi$
for $i$. We say that $\varphi$ respects unambiguous improvement\emph{s}
if it respects unambiguous improvements for each student $i\in I$. 
\end{defn}
Respect for improvements is essential in settings like ours where
it implies that students never want to intentionally decrease their
test scores and, in turn, their rankings. Similarly, it is also important
in cadet-branch matching where cadets can influence their priority
rankings directly. \emph{Sönmez} (2013) argues that cadets take perverse
steps to lower their priorities because the mechanism used by the
Reserve Officer Training Corps (ROTC) to match its cadets to branches
fails the respecting improvements property. 
\begin{thm}
The cumulative offer mechanism with respect to dynamic reserves choice
functions respects unambiguous improvements. 
\end{thm}

\paragraph{Proof. }

See Appendix 7.3. 

\section{Comparative Statics on Monotonic Capacity Transfer Schemes}

In this section, we first define a comparison criteria between two
monotone capacity transfer schemes. Consider a school $s\in S$ with
a given precedence sequence $f^{s}$ and target distribution $\overline{q}_{s}=(\overline{q}_{s}^{1},...,\overline{q}_{s}^{\lambda_{s}})$.
Let $q_{s}$ and $\widetilde{q}_{s}$ be two monotone capacity transfer
schemes: given a vector of unused slots from group of slots $1$ to
$j-1$, $(r_{1},...,r_{j-1})\in\mathbb{Z}_{+}^{j-1}$, the dynamic
capacity of the $j^{th}$ group under capacity transfer schemes $q_{s}$
and $\widetilde{q}_{s}$ are $q_{s}^{j}=q_{s}^{j}(r_{1},...,r_{j-1})$
and $\widetilde{q}_{s}^{j}=\widetilde{q}_{s}^{j}(r_{1},...,r_{j-1})$,
respectively, for all $j\geq2$ and, $q_{s}^{1}=\widetilde{q}_{s}^{1}=\overline{q}_{s}^{1}$.

Let $q_{s}$ and $\widetilde{q}_{s}$ be two monotone capacity transfer
schemes that are compatible with the precedence sequence $f^{s}$
and target capacity vector $\overline{q}_{s}$ of school $s\in S$.
We say that the monotone capacity transfer scheme $\widetilde{q}^{s}$
is $\mathbf{more}$ $\mathbf{flexible}$ than the monotone capacity
transfer scheme $q^{s}$ if
\begin{enumerate}
\item there exists $l\in\{2,...,\lambda_{s}\}$ and $(\hat{r}_{1},...,\hat{r}_{l-1})\in\mathbb{Z}_{+}^{l-1}$
such that $\widetilde{q}_{s}^{l}(\hat{r}_{1},...,\hat{r}_{l-1})>q_{s}^{l}(\hat{r}_{1},...,\hat{r}_{l-1})$,
and 
\item for all $j\in\{2,...,\lambda_{s}\}$ and $(r_{1},...,r_{j-1})\in\mathbb{Z}_{+}^{j-1}$,
if $j\neq l$ or $(r_{1},...,r_{j-1})\neq(\hat{r},...,\hat{r}_{l-1})$,
then $\widetilde{q}_{s}^{j}(r_{1},...,r_{j-1})\geq q_{s}^{j}(r_{1},...,r_{j-1})$. 
\end{enumerate}
The definition states that one monotonic capacity transfer scheme
is more flexible than another if it transfers at least as many vacant
slots as the other at every instance (i.e., the vectors of the number
of unused slots). There must also be an instance where the first one
transfers strictly more vacant slots than the second one to the next
group of slots according to the precedence sequence. Also, both of
the monotonic capacity transfer schemes take the capacity of the first
group of slots with respect to the precedence sequence equal to its
target capacity. Holding all else constant, when the capacity transfer
scheme becomes more flexible, it defines a particular choice function
expansion.\footnote{The type of choice function expansion here is different than the one
\emph{Chambers and Yenmez} (2017)\emph{ }define\emph{. }Their notion
of expansion is in the sense of set inclusion while ours is not. They
say that a choice function $C^{'}$ is an expansion of another choice
function $C$ if for every offer set $Y$, $C(Y)\subseteq C^{'}(Y)$.
According to the expansion via a more flexible capacity transfer scheme,
when a choice function $C$ expands to $C^{'}$ it is possible to
have $C(Y)\nsubseteq C^{'}(Y)$ for some $Y$. }

Expanding the overall choice function of a single school leads to
Pareto improvement for students under the COM.\footnote{This result does not contradict the findings of \emph{Alva and Manjunath
}(2019), because increasing flexibility of the capacity transfers
changes the choice functions, and therefore the set of contracts that
are feasible in their context. Theorem 4 achieves the improvement
by considering a dominating mechanism that is infeasible under the
original transfer scheme. } 
\begin{thm}
Let $C=(C^{s_{1}},...,C^{s_{m}})$ be the profile of schools' overall
choice functions. Fix a school $s\in S$. Suppose that $\widetilde{C}^{s}$
takes a capacity transfer scheme that is more flexible than that of
$C^{s}$, holding all else constant. Then, the outcome of the cumulative
offer mechanism with respect to $(\widetilde{C}^{s},C_{-s})$ weakly
Pareto dominates the outcome of the cumulative offer mechanism with
respect to $C$. 
\end{thm}

\paragraph{Proof.}

See Appendix 7.3. 

Theorem 4 is of particular importance because it indicates that increasing
the transferability of capacity from low-demand to high-demand groups
leads to strategy-proof Pareto improvement with the cumulative offer
algorithm. This result provides a normative foundation for recommending
a more flexible interpretation of type-specific quotas.  This result
establishes that to maximize students' welfare, schools' choice functions
should be expanded as much as possible. 

It is important to note that when more than one school's capacity
transfer scheme become more flexible, a simple iteration of Theorem
4, one school at a time, ensures (weak) Pareto improvement. Therefore,
a more flexible capacity transfer profile of schools implies that
the COM with the new capacity transfer scheme (weakly) Pareto improves
the original transfer scheme. 

\section{Relationship Between Slot-specific Priorities and Dynamic Reserves
Choice Rules}

In this section, we investigate the relationship between the families
of slot-specific priorities choice rules and dynamic reserves choice
rules. To do so, we first describe slot-specific priorities choice
rules. 

Each school $s\in S$ has a set of slots $\mathcal{B}_{s}$. Each
slot can be assigned at most one contract in $X_{s}$. Slots $b\in\mathcal{B}_{s}$
have linear priority orders $\pi^{b}$ over contracts in $X_{s}$.
Each slot $b$ ranks a null contract $\emptyset_{b}$ that represents
remaining unassigned. Schools $s\in S$ may be assigned as many as
$\mid\mathcal{B}_{s}\mid$ contracts from an offer set $Y\subseteq X$---one
for each slot in $\mathcal{B}_{s}$--- but may hold no more than
one contract with a given student. The slots in $\mathcal{B}_{s}$
are ordered according to a linear order of precedence $\triangleright^{s}$.
We denote $\mathcal{B}_{s}\equiv\{b_{1},...,b_{\overline{q}_{s}}\}$
with $\mid\mathcal{B}_{s}\mid=\overline{q}_{s}$. The interpretation
of $\triangleright^{s}$ is that if $b_{l}\triangleright^{s}b_{l+1}$,
then---whenever possible---school $s$ fills slot $b_{l}$ before
filling slot $b_{l+1}$. Formally, the choice $C^{s}(Y)$ of a school
$s\in S$ from contract set $Y\subseteq X$ is defined as follows: 
\begin{itemize}
\item First, slot $b_{1}$ is assigned the contract $y_{1}$ that is $\pi^{b_{1}}$-maximal
among contracts in $Y$. 
\item Then, slot $b_{2}$ is assigned the contract $y_{2}$ that is $\pi^{b_{2}}$-maximal
among contracts in the set $Y\setminus Y_{\mathbf{i}(y_{1})}$ of
contracts in $Y$ with agents other than $\mathbf{i}(y_{1})$. 
\item This process continues in sequence, with each slot $b_{l}$ being
assigned to the contract $y_{l}$ that is $\pi^{b_{l}}$-maximal among
contracts in the set $Y\setminus Y_{\mathbf{i}(\{y_{1},...,y_{l-1}\})}$. 
\end{itemize}
If no contract is assigned to a slot $b_{l}\in\mathcal{B}_{s}$ in
the computation of $C^{s}(Y)$, then $b_{l}$ is assigned the null
contract $\emptyset_{b_{l}}$. 

We first give an example of a dynamic reserves choice rule that cannot
be generated by a slot-specific priorities choice rule. 
\begin{example}
Consider $I=\{i,j,k,l\}$, $S=\{s\}$ with $q_{s}=2$, and $\Theta=\{t_{1},t_{2},t_{3}\}$.
Student $i$ only has type $t_{1}$ and a single contract $x_{1}$.
Student $j$ only has type $t_{2}$ and a single contract $y_{2}$.
Student $k$ has types $t_{2}$ and $t_{3}$, and two contracts related
to these types $z_{2}$ and $z_{3}$, respectively. Finally, student
$l$ has types $t_{1}$ and $t_{3}$, and two contracts related to
these types $w_{1}$ and $w_{3}$, respectively. The set of contracts
for this problem is $X=\{x_{1},y_{2},z_{2},z_{3},w_{1},w_{3}\}$.
Students are ordered with respect to their exam scores from highest
to lowest as follows: $i-j-k-l$. 
\end{example}
The school reserves the first seat for type $t_{1}$, and the second
seat for type $t_{2}$. If either the first seat or the second seat
cannot be filled with the students they are reserved for, they are
filled with a type $t_{3}$ student(s). The precedence order is such
that the first seat is filled first with a type $t_{1}$ student if
possible, and then the second seat is filled with a type $t_{2}$
student, if possible. If any of these seats cannot be filled with
the intended student types, all of the vacant seats are filled with
type $t_{3}$ students at the very end, if possible. 

We can represent the distributional objective described above by capacity-transfers
as follows: Initially $\overline{q}_{t_{1}}=\overline{q}_{t_{2}}=1$
and $\overline{q}_{t_{3}}=0$. The dynamic capacity of the third seat
is given by $q_{t_{3}}=r_{1}+r_{2}$, where $r_{1},r_{2}\in\{0,1\}$.
Some of the choice situations under the capacity-transfer described
above are given below: \bigskip{}

\[
\begin{array}{cc}
Y & C(Y)\\
\{x_{1},y_{2},z_{2},z_{3},w_{1},w_{3}\} & \{x_{1},y_{2}\}\\
\{y_{2},z_{2},z_{3}\} & \{y_{2},z_{3}\}\\
\{x_{1},z_{2},z_{3}\} & \{x_{1},z_{2}\}\\
\{y_{2},w_{1},w_{3}\} & \{y_{2},w_{1}\}\\
\{x_{1},w_{1},w_{3}\} & \{x_{1},w_{3}\}\\
\{z_{2},z_{3}\} & \{z_{2}\}\\
\{w_{1},w_{3}\} & \{w_{1}\}
\end{array}
\]

In order to implement the choices above with slot-specific priorities,
we need to find a strict ranking of the contracts in $X$ for both
of the slots. Note that $\{x_{1},y_{2}\}$ is chosen from $\{x_{1},y_{2},z_{2},z_{3},w_{1},w_{2}\}$.
Then, $x_{1}$ must be chosen for one of the slots and $y_{2}$ must
be chosen for the other. There are two cases to consider. 

\paragraph*{Case 1: }

$x_{1}$ is chosen from slot 1 and $y_{2}$ is chosen from slot 2.
Then, $x_{1}$ is the highest priority contract in slot 1. We have
$C(\{x_{1},z_{2},z_{3}\})=\{x_{1},z_{2}\}$. Then, $z_{2}$ must have
higher priority than $z_{3}$ in the strict priority ranking of slot
2 because $x_{1}$ will be chosen from the first slot. Notice that
both $z_{2}$ and $z_{3}$ must have lower priority than $y_{2}$
in the strict ranking of slot 2. Also, since $C(\{y_{2},z_{2},z_{3}\})=\{y_{2},z_{3}\}$,
then it must be the case that $z_{3}$ has higher priority than $z_{2}$
in the strict priority of the first slot. Notice that $z_{3}$ cannot
be chosen from the second slot as $z_{2}$ has higher priority. However,
$C(\{z_{2},z_{3}\})=\{z_{2}\}$. This is a contradiction. 

\paragraph*{Case 2: }

$y_{2}$ is chosen from slot 1 and $x_{1}$ is chosen from slot 2.
Then, $y_{2}$ has the highest priority in slot 1. We have $C(\{y_{2},w_{1},w_{3}\})=\{y_{2},w_{1}\}$.
Therefore, in the ranking of slot 2, $w_{1}$ must have higher priority
than $w_{3}$. Also, since $C(\{x_{1},w_{1},w_{3}\})=\{x_{1},w_{3}\}$,
it follows that in the ranking of slot 1 $w_{3}$ must have higher
priority than $w_{1}$. This is because $w_{3}$ cannot be chosen
from slot 2 as it has a lower priority than $w_{1}$ there. However,
$C(\{w_{1},w_{3}\})=\{w_{1}\}$. This is a contradiction. 

Hence, we cannot find a strict rankings of the contracts in $X$ for
these two slots that generate the dynamic reserves choice rule defined
above. 

Our last result states that the family of dynamic reserves choice
rules nests the family of slot-specific priorities choice rules. 

\begin{thm}
Every slot-specific priorities choice rule can be generated by a dynamic
reserves choice rule. 
\end{thm}

\paragraph{Proof.}

See Appendix 7.3. 

\section{Conclusion}

This paper studies a school choice problem with distributional objectives
where students care about both the school they are matched with as
well as the category through which they are admitted. Each school
can be thought of as union of different groups of slots, where each
group is associated with exactly one category. Schools have target
distributions over their groups of slots in the form of reserves.
If these reserves are considered to be hard bounds, then some slots
will remain empty in instances where demand for particular categories
is less than their target capacities. To overcome this problem and
to increase efficiency, we design a family of dynamic reserves choice
functions. We do so by allowing monotonic capacity transfers across
groups of slots when one or more of the groups is not able to fill
to its target capacity. The capacity transfer scheme is exogenously
given for each school and governs the dynamic capacities of groups,
each of which has a q-responsive sub-choice function. The overall
choice function of a school can be thought of as the union of choices
with these sub-choice functions of its groups. 

We offer the COM with respect to dynamic reserves choice functions
as an allocation rule. We show that the COM is stable and strategy-proof
in our framework. Moreover, the COM respects improvements. We introduce
a comparison criteria between two monotonic capacity transfer schemes.
If a monotone capacity transfer scheme transfers at least as many
vacancies in every contingency compared to another monotone capacity
transfer scheme, we say that the first is more flexible than the second.
We show that when capacity transfer scheme of a school becomes more
flexible, while other school choice functions remain unchanged, the
outcome of the COM under the modified profile of choice functions
Pareto dominates the outcome of the COM under the original profile.
This result is the main message of our paper, as it describes a strategy-proof
Pareto improvement by making capacity transfers more flexible. 

\section{APPENDICES }

\subsection{Formal Description of the Cumulative Offer Process}

\paragraph*{Cumulative Offer Process (COP): }

Consider the outcome the COM as denoted by $\Phi^{\Gamma}\left(P,C\right)$.
For any preference profile $P$ of students, profile of choice functions
of schools $C$, and an ordering $\Gamma$ of the elements of $X$,
the outcome is determined by the \emph{COP  with respect to $\Gamma$,
$P$ }and\emph{ $C$ }as follows:

\paragraph*{Step 0: }

Initialize the set of contracts \emph{available }to the schools as
$A^{0}=\emptyset$. 

\paragraph*{Step $t\protect\geq1$:}

Consider the set 
\[
U^{t}\equiv\left\{ x\in X\setminus A^{t-1}:\mathbf{i}(x)\notin\mathbf{i}\left(C^{S}(A^{t-1})\right)\;and\;\nexists z\in\left(X_{\mathbf{i}(x)}\setminus A^{t-1}\right)\cup\left\{ \emptyset\right\} \;such\;that\;zP^{\mathbf{i}(x)}x\right\} .
\]

If $U^{t}$ is empty, then the algorithm terminates and the outcome
is given by $C^{S}(A^{t-1})$.\footnote{We denote by $C^{S}(Y)\equiv\cup_{s\in S}C^{s}(Y)$ the set of contracts
chosen by the set of schools from a set of contracts $Y\subseteq X$.} Otherwise, letting $y^{t}$ be the highest-ranked element of $U^{t}$
according to $\Gamma$, we say that $y^{t}$ is \emph{proposed }and
set $A^{t}=A^{t-1}\cup\{y^{t}\}$ and proceed to step $t+1$. 

A COP begins with no contracts available to the schools (i.e., $A^{0}=\emptyset$).
Then, at each step $t$, we construct $U^{t}$, the set of contracts
that (1) have not yet been proposed, (2) are not associated to students
with contracts chosen by schools from the currently available set
of contracts, and (3) are both acceptable and the most-preferred by
their associated students among all contracts not yet proposed. If
$U^{t}$ is empty, then every student $i$ either has some associated
contract chosen by some school, i.e., $i\in\mathbf{i}\left(C^{S}(A^{t-1})\right)$,
or has no acceptable contracts left to propose, and so the COP  ends.
Otherwise, the contract in $U^{t}$ that is highest-ranked according
to $\Gamma$ is proposed by its associated student, and the process
proceeds to the next step. Note that at some step this process must
end as the number of contracts is finite. 

Letting $T$ denote the last step of the COP, we call $A^{T}$ the
set of contracts observed in the COP  with respect to $\Gamma$, $P$,
and $C$. 

\subsection{Substitutable Completion of Dynamic Reserves Choice Functions}
\begin{defn}
A choice function $C^{s}(\cdot)$ satisfies the \textbf{irrelevance
of rejected contracts} (IRC) condition if for all $Y\subset X$, for
all $z\in X\setminus Y$, and $z\notin C^{s}\left(Y\cup\left\{ z\right\} \right)\Longrightarrow C^{s}\left(Y\right)=C^{s}\left(Y\cup\left\{ z\right\} \right)$.
\end{defn}
\emph{Hatfield and Milgrom} (2005) introduce the substitutability
condition, which generalizes the earlier gross substitutes\emph{ }condition
of \emph{Kelso and Crawford} (1982).
\begin{defn}
A choice function $C^{s}(\cdot)$ satisfies \textbf{substitutability
}if for all $z,\,z'\in X$, and $Y\subseteq X$, $z\notin C^{s}\left(Y\cup\left\{ z\right\} \right)\Longrightarrow z\notin C^{s}\left(Y\cup\left\{ z,z'\right\} \right)$.
\end{defn}
\begin{defn}
A choice function $C^{s}\left(\cdot\right)$ satisfies the \textbf{law
of aggregate demand }(LAD) if $Y\subseteq Y'\Longrightarrow\mid C^{s}\left(Y\right)\mid\,\leq\,\mid C^{s}\left(Y'\right)\mid$.
\end{defn}
The following definitions are from \emph{Hatfield and Kominers} (2019).
A \emph{completion }of a many-to-one choice function $C^{s}(\cdot)$
of school $s\in S$ is a choice function $\overline{C}^{s}(\cdot)$,
such that for all $Y\subseteq X$, either $\overline{C}^{s}(Y)=C^{s}(Y)$
or there exists a distinct $z,z^{'}\in\overline{C}^{s}(Y)$ such that
$i(z)=i(z^{'})$. If a choice function $C^{s}(\cdot)$ has a completion
that satisfies the substitutability and \emph{IRC} condition, then
we say that $C^{s}(\cdot)$ is \emph{substitutably completable. }If
every choice function in a profile $C=(C^{s}(\cdot))_{s\in S}$ is
substitutably completable, then we say that $C$ is \emph{substitutably
completable.}

Let $C^{s}(\cdot,f^{s},q_{s})$ be a dynamic reserve choice function
given the precedence sequence $f^{s}$ and the capacity transfer scheme
$q_{s}$. We define a related choice function $\overline{C}^{s}(\cdot,f^{s},q_{s})$.
Given a set of contracts $Y\subseteq X$, $\overline{C}^{s}(Y,f^{s},q_{s})$
denotes the set of chosen contracts from set $Y$ and is determined
as follows: 
\begin{itemize}
\item Given $\overline{q}_{s}^{1}$ and $Y=Y^{0}\subseteq X$, let $Y_{1}\equiv c_{1}^{s}(Y^{0},\overline{q}_{s}^{1},f^{s}(1))$
be the set of chosen contracts with privilege $f^{s}(1)$ from $Y^{0}$.
Let $r_{1}=\overline{q}_{s}^{1}-\mid Y_{1}\mid$ be the number of
vacant slots. The set of remaining contracts is then $Y^{1}=Y^{0}\setminus Y_{1}$. 
\item In general, let $Y_{k}=c_{k}^{s}(Y^{k-1},q_{s}^{k},f^{s}(k))$ be
the set of chosen contracts with privilege $f^{s}(k)$ from the set
of available contracts $Y^{k-1}$ , where $q_{s}^{k}=q_{s}^{k}(r_{1},...,r_{k-1})$
is the dynamic capacity of group of slots $k$ as a function of the
vector of the number of unfilled slots $(r_{1},...,r_{k-1})$. Let
$r_{k}=q_{s}^{k}-\mid Y_{k}\mid$ be the number of vacant slots. The
set of remaining contracts is then $Y^{k}=Y^{k-1}\setminus Y_{k-1}$. 
\item Given $Y=Y^{0}\subseteq X$ and the capacity of the first group of
slots $\overline{q}_{s}^{1}$ , we define $\overline{C}^{s}(Y,f^{s},q_{s})=c_{1}^{s}(Y^{0},\overline{q}_{s}^{1},f^{s}(1))\cup(\stackrel[k=2]{\lambda_{s}}{\cup}c_{k}^{s}(Y^{k-1},q_{s}^{k}(r_{1},...,r_{k-1}),f^{s}(k)))$. 
\end{itemize}
The difference between $C^{s}(\cdot)$ and $\overline{C}^{s}(\cdot)$
is as follows: In the computation of $C^{s}(\cdot)$, if a contract
of a student is chosen by some group of slots then his/her other contracts
are removed for the rest of the choice procedure. However, in the
computation of $\overline{C}^{s}(\cdot)$ this is not the case. According
to the choice procedure $\overline{C}^{s}(\cdot)$, if a contract
of a student is chosen, say, by group of slots $k$, then his/her
other contracts will still be available for the following groups of
slots. 

The following proposition shows that $\overline{C}^{s}(\cdot)$ defined
above is the completion of the dynamic reserves choice function  $C^{s}(\cdot)$. 
\begin{prop}
$\overline{C}^{s}(\cdot)$ is a completion of $C^{s}(\cdot)$. 
\end{prop}

\paragraph{Proof. }

Let $f^{s}$ and $q_{s}$ be the precedence sequence and capacity
transfer scheme of school $s\in S$, respectively. Take an offer set
$Y=Y^{0}\subseteq X$ and assume there is no pair of contracts $z,z^{'}\in Y^{0}$
such that $i(z)=i(z^{'})$ and $z,z^{'}\in\overline{C}^{s}(Y,f^{s},q_{s})$.
We want to show that 
\[
\overline{C}^{s}(Y,f^{s},q_{s})=C^{s}(Y,f^{s},q_{s}).
\]

Let $Y_{j}$ be the set of contracts chosen by group of slots $j$
and let $Y^{j}$ be the set of contracts that remains in the choice
procedure after group $j$ selects according to dynamic reserve choice
function $C(\cdot)$. Similarly, let $\overline{Y}_{j}$ be the set
of contracts chosen by group of slots $j$ and let $\overline{Y}^{j}$
be the set of contracts that remains in the choice procedure after
group $j$ selects according to the completion $\overline{C}(\cdot)$.
Notice that $Y^{0}=\overline{Y}^{0}$. Let $r_{j}$ and $\overline{r}_{j}$
be the number of vacant slots in group of slots $j$ in the choice
procedures $C^{s}(Y,f^{s},q_{s})$ and $\overline{C}^{s}(Y,f^{s},q_{s})$,
respectively. Also, let $q_{s}^{j}(r_{1},...,r_{j-1})$ and $\overline{q}_{s}^{j}(\overline{r}_{1},...,\overline{r}_{j-1})$
denote the dynamic capacities of group of slots $j$ under choice
procedures $C^{s}(Y,f^{s},q_{s})$ and $\overline{C}^{s}(Y,f^{s},q_{s})$,
respectively. 

Given $\bar{q}_{s}^{1}$ and $Y^{0}=\overline{Y}^{0}$, we have $\overline{Y}_{1}=c_{1}^{s}(Y^{0},\bar{q}_{s}^{1},f^{s}(1))=Y_{1}$
by the construction of $\overline{C}^{s}$. Moreover, $\overline{r}_{1}=r_{1}$
and $\overline{q}_{s}^{2}(\overline{r}_{1})=q_{s}^{2}(r_{1})$. 

Suppose that for all $j\in\{2,...,k-1\}$ we have $Y_{j}=\overline{Y}_{j}$.
We need to show that it holds for group of slots $k$, i.e., $Y_{k}=\overline{Y}_{k}$.
Since the chosen set is the same in every group from $1$ to $k-1$
under $C(\cdot)$ and $\overline{C}(\cdot)$, the number of remaining
slots in each group is the same as well. Then, the dynamic capacity
of the group of slots $k$ are the same under choice procedures $C^{s}(Y,f^{s},q_{s})$
and $\overline{C}^{s}(Y,f^{s},q_{s})$, i.e., $q_{s}^{k}(r_{1},...,r_{k-1})=\overline{q}_{s}^{k}(\overline{r}_{1},...,\overline{r}_{k-1})$.
Since there are no two contracts of an agent chosen by $\overline{C}^{s}(Y,f^{s},q_{s})$,
one can deduce that all of the remaining contracts of agents, whose
contracts were chosen by previous sub-choice functions, are rejected
by $c_{k}^{s}(\overline{Y}^{k-1},\overline{q}_{s}^{k}(\overline{r}_{1},...,\overline{r}_{k-1}),f^{s}(k))$.
Therefore, the IRC of the sub-choice function implies that
\[
c_{k}^{s}(\overline{Y}^{k-1},\overline{q}_{s}^{k}(\overline{r}_{1},...,\overline{r}_{k-1}),f^{s}(k))=c_{k}^{s}(Y^{k-1},q_{s}^{k}(r_{1},...,r_{k-1}),f^{s}(k)).
\]
Hence, we have $\overline{Y}_{k}=Y_{k}$, $\overline{r}_{k}=r_{k}$,
and $\overline{q}_{s}^{k+1}(\overline{r}_{1},...,\overline{r}_{k})=q_{s}^{k+1}(r_{1},...,r_{k})$. 

Since in each group of slots the same sets of contracts are chosen
by the dynamic reserve choice function and its completion, the result
follows. 

\begin{prop}
$\overline{C}^{s}(\cdot)$ satisfies the IRC.
\end{prop}

\paragraph{Proof. }

For any $Y\subseteq X$ such that $Y\neq\overline{C}^{s}(Y,f^{s},q_{s})$,
let $x$ be one of the rejected contracts, i.e., $x\in Y\setminus\overline{C}^{s}(Y,f^{s},q_{s})$.
To show that the IRC is satisfied, we need to prove that 
\[
\overline{C}^{s}(Y,f^{s},q_{s})=\overline{C}^{s}(Y\setminus\{x\},f^{s},q_{s}).
\]

Let $\tilde{Y}=Y\setminus\{x\}$. Let $(\overline{Y}_{j},\bar{r}_{j},\overline{Y}^{j})$
be the sequence of the set of chosen contracts, the number of vacant
slots, and the remaining set of contracts for group $j=1,...,\lambda_{s}$
from $Y$ under $\overline{C}(\cdot)$. Similarly, let $(\tilde{Y}_{j},\tilde{r}_{j},\tilde{Y}^{j})$
be the sequence of the set of chosen contracts, the number of vacant
slots, and the remaining set of contracts for group $j=1,...,\lambda_{s}$
from $\tilde{Y}$ under $\overline{C}(\cdot)$. 

For the first group of slots, since the sub-choice functions satisfy
the IRC, we have $\overline{Y}_{1}=\tilde{Y}_{1}$. Moreover, $\bar{r}_{1}=\tilde{r}_{1}$
and $\overline{Y}^{1}\setminus\{x\}=\tilde{Y}^{1}$. By induction,
for each $j=2,...,k-1$, assume that 

\[
\overline{Y}_{j}=\tilde{Y}_{j},\;\bar{r}_{j}=\tilde{r}_{j},\;and\;\overline{Y}^{j}\setminus\{x\}=\tilde{Y}^{j}.
\]

We need to show that the above equalities hold for $j=k$. Since ,
$x\notin\overline{C}^{s}(Y,f^{s},q_{s})$ and the sub-choice functions
satisfy the IRC condition we have 
\[
c_{k}^{s}(\overline{Y}^{k-1},q_{s}^{k}(\overline{r}_{1},...,\overline{r}_{k-1}),f^{s}(k))=c_{k}^{s}(\tilde{Y}^{k-1},q_{s}^{k}(\tilde{r}_{1},...,\tilde{r}_{k-1}),f^{s}(k)).
\]
The same set of contracts is chosen for group $k$ in the choice processes
beginning with $Y$ and $Y\cup\{x\}$, respectively. By our inductive
assumption that $\bar{r}_{j}=\tilde{r}_{j}$ for each $j=2,...,k-1$,
the dynamic capacity of group $k$ is the same under both choice processes.
The number of remaining slots is the same as well, i.e., $\bar{r}_{k}=\tilde{r}_{k}$.
Finally, we know that $x$ is chosen from the set $\tilde{Y}^{k-1}\cup\{x\}$,
then we have 
\[
\overline{Y}^{k}=\tilde{Y}^{k}\cup\{x\}.
\]

Since for all $j\in\{1,...,\lambda_{s}\}$, $\overline{Y}_{j}=\tilde{Y}_{j}$,
we have $\overline{C}^{s}(Y,f^{s},q_{s})=\overline{C}^{s}(\tilde{Y},f^{s},q_{s})$.
Hence, $\overline{C}^{s}(\cdot,f^{s},q_{s})$ satisfies the IRC. 

\begin{prop}
$\overline{C}^{s}(\cdot)$ satisfies the substitutability. 
\end{prop}

\paragraph{Proof. }

Consider an offer set $Y\subseteq X$ such that $Y\neq\overline{C}^{s}(Y,f^{s},q_{s})$.
Let $x$ be one of the rejected contracts, i.e., $x\in Y\setminus\overline{C}^{s}(Y,f^{s},q_{s})$,
and let $z$ be an arbitrary contract in $X\setminus Y$. To show
substitutability, we need to show that 
\[
x\notin\overline{C}^{s}(Y\cup\{z\},f^{s},q_{s}).
\]
 Consider $\tilde{Y}=Y\cup\{z\}$. Let $(Y_{j},r_{j},Y^{j})$ be the
sequence of the set chosen contracts, the number of vacant slots,
and the set of remaining contracts for group of slots $j=1,...,\lambda_{s}$
from $Y$ under $\overline{C}(\cdot)$. Similarly, let $(\tilde{Y}_{j},\tilde{r}_{j},\tilde{Y}^{j})$
be the sequence of the set chosen contracts, the number of vacant
slots, and the set of remaining contracts for group of slots $j=1,...,\lambda_{s}$
from $\tilde{Y}$ under $\overline{C}(\cdot)$. There are two cases
to consider:

\paragraph*{Case 1}

$z\in\tilde{Y}\setminus\overline{C}^{s}(\tilde{Y},f^{s},q_{s})$.

In this case, the IRC of $\overline{C}^{s}$ implies $\overline{C}^{s}(\tilde{Y},f^{s},q_{s})=\overline{C}^{s}(Y,f^{s},q_{s})$.
Therefore, $x\notin\overline{C}^{s}(\tilde{Y},f^{s},q_{s})$. 

\paragraph*{Case 2}

$z\in\overline{C}^{s}(\tilde{Y},f^{s},q_{s})$. 

Let $j$ be the group of slots such that $z\in\tilde{Y}_{j}$. By
the IRC of sub-choice functions, $x\notin\tilde{Y}_{j}=Y_{j}$, for
all $j^{'}=1,...,j-1$. Moreover, $\tilde{Y}^{j^{'}-1}=Y^{j^{'}-1}\cup\{z\}$
and $\tilde{r}_{j^{'}}=r_{j^{'}}$, for all $j^{'}=1,...,j-1$. 

First note that the dynamic capacity of group $j$ is the same under
choice procedures beginning with $Y=Y^{0}$ and $Y\cup\{z\}=\tilde{Y}^{0}$,
respectively. This is because the number of unused slots from groups
$1$ to $j-1$ are the same under the two choice procedures. We know
that $z$ is chosen exactly at group $j$ in the process beginning
with $\tilde{Y}^{0}$. There are two cases here: 

\paragraph{(a) }

The dynamic capacity of group $j$ is exhausted in the process beginning
with $Y^{0}$. In this case, by choosing $z$ from $\tilde{Y}^{0}$
another contract, we say that say $y\in\tilde{Y}^{0}$ is rejected
even though $y$ was chosen at group $j$ in the process beginning
with $Y^{0}$. 

\paragraph{(b) }

The dynamic capacity of group $j$ is $\mathbf{not}$ exhausted in
the choice process beginning with $Y^{0}$. In this case, $z$ is
chosen at group $j$ in the process beginning with $\tilde{Y}^{0}$
without rejecting any contract that was chosen in the process beginning
with $Y^{0}$ at group $j$. 

In the case of $(a)$, 
\[
\mid c_{j}^{s}(Y^{j-1},q_{s}^{j}(r_{1},...,r_{j-1}),f^{s}(j))\mid=q_{s}^{j}(r_{1},...,r_{j-1})
\]
 and 
\[
z\in c_{j}^{s}(\tilde{Y}^{j-1},q_{s}^{j}(r_{1},...,r_{j-1}),f^{s}(j))
\]
 implies that there exists a contract $y$ such that 
\[
y\in c_{j}^{s}(Y^{j-1},q_{s}^{j}(r_{1},...,r_{j-1}),f^{s}(j))\setminus c_{j}^{s}(\tilde{Y}^{j-1},q_{s}^{j}(\tilde{r}_{1},...,\tilde{r}_{j-1}),f^{s}(j)).
\]
 This implies that $\tilde{Y}^{j}=Y^{j}\cup\{y\}$. Since the capacity
of group $j$ is exhausted under both choice processes, the number
of vacant slots for group $j$ will be $0$ in both choice processes.
Thus, the capacity will be the same for group $j+1$ under both. 

Notice that 
\[
x\notin Y_{j}\;\Longrightarrow x\notin\tilde{Y}_{j}
\]
 because 
\[
c_{j}^{s}(Y^{j-1},q_{s}^{j}(r_{1},...,r_{j-1}),f^{s}(j))\cup\{z\}\setminus\{y\}=c_{j}^{s}(\tilde{Y}^{j-1},q_{s}^{j}(\tilde{r}_{1},...,\tilde{r}_{j-1}),f^{s}(j)).
\]

In case $(b)$, we have 
\[
\mid c_{j}^{s}(Y^{j-1},q_{s}^{j}(r_{1},...,r_{j-1}),f^{s}(j))\mid<q_{s}^{j}(r_{1},...,r_{j-1}).
\]
 Hence, $r_{j}>0$. Then, since the sub-choice functions are responsive,
we have 
\[
c_{j}^{s}(\tilde{Y}^{j-1},q_{s}^{j}(\tilde{r}_{1},...,\tilde{r}_{j-1}),f^{s}(j))=\{z\}\cup c_{j}^{s}(Y^{j-1},q_{s}^{j}(r_{1},...,r_{j-1}),f^{s}(j)).
\]

Therefore, 
\[
x\notin Y_{j}\;\Longrightarrow\;x\notin\tilde{Y}_{j}.
\]
We also have $r_{j}=\tilde{r}_{j}+1$. Moreover, the set of remaining
contracts under both choice processes will be the same, i.e., $\tilde{Y}^{j}=Y^{j}$.
The facts $r_{j^{'}}=\tilde{r}_{j^{'}}$ for all $j^{'}=1,...,j-1$
and $r_{j}=\tilde{r}_{j}+1$ implies---by the monotonicity of capacity
transfer schemes---that either 
\[
q_{s}^{j+1}(r_{1},...,r_{j})=q_{s}^{j+1}(\tilde{r}_{1},...,\tilde{r}_{j})
\]
 or 
\[
q_{s}^{j+1}(r_{1},...,r_{j})=1+q_{s}^{j+1}(\tilde{r}_{1},...,\tilde{r}_{j})
\]
 hold. 

Suppose now that for all $\gamma=j,...,k-1$ we have that either 
\[
\left[\tilde{Y}^{\gamma}=Y^{\gamma}\cup\{\widetilde{y}\}\;for\;some\;\widetilde{y}\;and\;q_{s}^{\gamma+1}(\tilde{r}_{1},...,\tilde{r}_{\gamma})=q_{s}^{\gamma+1}(r_{1},...,r_{\gamma})\right]
\]
 or 
\[
\left[\tilde{Y}^{\gamma}=Y^{\gamma}\;and\;q_{s}^{\gamma+1}(\tilde{r}_{1},...,\tilde{r}_{\gamma})\leq q_{s}^{\gamma+1}(r_{1},...,r_{\gamma})\leq1+q_{s}^{\gamma+1}(\tilde{r}_{1},...,\tilde{r}_{\gamma})\right].
\]

We have already shown that it holds for $\gamma=j$ and we will now
show that it also holds for $\gamma=k$. 

We will first analyze the former case. By inductive assumption, we
have $\tilde{Y}^{k-1}=Y^{k-1}\cup\{\widetilde{y}\}$ for some contract
$\widetilde{y}$. If $\widetilde{y}$ is not chosen from the set $\widetilde{Y}^{k-1}$
then exactly the same set of contracts will be chosen from $Y^{k-1}$
and $\tilde{Y}^{k-1}$ since the capacities of group $k$ are the
same under both choice processes and the sub-choice function satisfies
the IRC condition. Then, we will have $\tilde{Y}^{k}=Y^{k}\cup\{\widetilde{y}\}$.
Moreover, since the number of vacant slots at group $k$ will be the
same under both processes, we will have $q_{s}^{k+1}(r_{1},...,r_{j})=q_{s}^{k+1}(\tilde{r}_{1},...,\tilde{r}_{j})$.
If $\widetilde{y}$ is chosen from the set $\tilde{Y}^{k-1}$, we
have two sub-cases, depending on if the dynamic capacity of group
$k$ is exhausted under the choice process beginning with $Y^{0}$.
If it is not exhausted, then we will have 
\[
c_{k}^{s}(\tilde{Y}^{k-1},q_{s}^{k}(\tilde{r}_{1},...,\tilde{r}_{k-1}),f^{s}(k))=\{\widetilde{y}\}\cup c_{k}^{s}(Y^{k-1},q_{s}^{k}(r_{1},...,r_{k-1}),f^{s}(k)),
\]
 which implies that $\tilde{Y}^{k}=Y^{k}$. Moreover, we will have
$r_{k}=\tilde{r}_{k}+1$. The monotonicity of capacity transfer scheme
implies that 
\[
q_{s}^{k+1}(\tilde{r}_{1},...,\tilde{r}_{k})\leq q_{s}^{k+1}(r_{1},...,r_{k})\leq1+q_{s}^{k+1}(\tilde{r}_{1},...,\tilde{r}_{k}).
\]

The first inequality follows from the fact that $\tilde{r}_{i}\leq r_{i}$
for all $i=1,...,k$. The second inequality follows from the second
condition of the monotonicity of the capacity transfer schemes. 

On the other hand, if the dynamic capacity of group $k$ is exhausted
in the choice procedure beginning with $Y^{0}$, then choosing $\widetilde{y}$
from the set $\tilde{Y}^{k-1}$ implies that there exists a contract
$\overline{y}$ that is chosen from $Y^{k-1}$ but rejected from $\tilde{Y}^{k-1}$.
Then, we will have $\tilde{Y}^{k}=Y^{k}\cup\{\overline{y}\}$ since
the sub-choice function is q-responsive and group $k$'s capacities
are the same under both choice processes. In this case, we will have
$r_{k}=\tilde{r}_{k}=0$. Since $\tilde{r}_{i}\leq r_{i}$ for all
$i=1,...,k$, we will have $q_{s}^{k+1}(\tilde{r}_{1},...,\tilde{r}_{k})\leq q_{s}^{k+1}(r_{1},...,r_{k})$
from the first condition of the monotonicity of the capacity transfer
scheme. Since $q_{s}^{k}(\tilde{r}_{1},...,\tilde{r}_{k-1})=q_{s}^{k}(r_{1},...,r_{k-1})$
and $\tilde{r}_{k}=r_{k}$, we will have $q_{s}^{k+1}(\tilde{r}_{1},...,\tilde{r}_{k})\geq q_{s}^{k+1}(r_{1},...,r_{k})$
by the second condition of the monotonicity of capacity transfer schemes.\footnote{In the second condition of the monotonicity of the capacity transfer
schemes, if the number of vacant slots is written as the dynamic capacity
of the group minus the number of chosen contracts then we will have
the following: the dynamic capacity of the group $k+1$ in the choice
process beginning with $Y$ minus the dynamic capacity of the group
$k+1$ in the choice process beginning with $Y\cup\{z\}=\tilde{Y}^{0}$
must be less than or equal to the summation of the difference of the
number of chosen contracts from group $1$ to group $k$, which is
0 in this specific case. } 

We will now analyze the latter case in which we have $\tilde{Y}^{k-1}=Y^{k-1}$
and either $q_{s}^{k}(r_{1},...,r_{k-1})=q_{s}^{k}(\tilde{r}_{1},...,\tilde{r}_{k-1})$
or $q_{s}^{k}(r_{1},...,r_{k-1})=1+q_{s}^{k}(\tilde{r}_{1},...,\tilde{r}_{k-1})$. 

If $q_{s}^{k}(r_{1},...,r_{k-1})=q_{s}^{k}(\tilde{r}_{1},...,\tilde{r}_{k-1})$,
then given that $\tilde{Y}^{k-1}=Y^{k-1}$, we will have $\tilde{Y}^{k}=Y^{k}$.
This also implies $r_{k}=\tilde{r}_{k}$. Moreover, we obtain $q_{s}^{k+1}(r_{1},...,r_{k})=q_{s}^{k+1}(\tilde{r}_{1},...,\tilde{r}_{k})$
by the monotonicity of capacity transfer scheme. Note that $\tilde{r}_{i}\leq r_{i}$
for all $i=1,...,k$ implies $q_{s}^{k+1}(r_{1},...,r_{k})\geq q_{s}^{k+1}(\tilde{r}_{1},...,\tilde{r}_{k})$
by the first condition of the monotonicity of capacity transfers.
The second condition of the monotonicity of capacity transfers implies
$q_{s}^{k+1}(r_{1},...,r_{k})\leq q_{s}^{k+1}(\tilde{r}_{1},...,\tilde{r}_{k})$. 

If $q_{s}^{k}(r_{1},...,r_{k-1})=1+q_{s}^{k}(\tilde{r}_{1},...,\tilde{r}_{k-1})$,
then given $\tilde{Y}^{k-1}=Y^{k-1}$, we have two sub-cases here. 

\paragraph{Sub-case 1. }

If 
\[
c_{k}^{s}(\tilde{Y}^{k-1},q_{s}^{k}(\tilde{r}_{1},...,\tilde{r}_{k-1}),f^{s}(k))=c_{k}^{s}(Y^{k-1},q_{s}^{k}(r_{1},...,r_{k-1}),f^{s}(k)),
\]
 then we will have $\tilde{Y}^{k}=Y^{k}$. Also, the monotonicity
of capacity transfer scheme implies that 
\[
q_{s}^{k+1}(\tilde{r}_{1},...,\tilde{r}_{k})\leq q_{s}^{k+1}(r_{1},...,r_{k})\leq1+q_{s}^{k+1}(\tilde{r}_{1},...,\tilde{r}_{k}).
\]

\paragraph{Sub-case 2. }

If 
\[
c_{k}^{s}(\tilde{Y}^{k-1},q_{s}^{k}(\tilde{r}_{1},...,\tilde{r}_{k-1}),f^{s}(k))\cup\{y^{*}\}=c_{k}^{s}(Y^{k-1},q_{s}^{k}(r_{1},...,r_{k-1}),f^{s}(k))
\]

for some $y^{*}$, then we will have $\tilde{Y}^{k}=Y^{k}\cup\{y^{*}\}$.
Moreover, the monotonicity of capacity transfer schemes in this case
implies that 
\[
q_{s}^{k+1}(r_{1},...,r_{k})=q_{s}^{k+1}(\tilde{r}_{1},...,\tilde{r}_{k}).
\]
This is because given $\tilde{r}_{i}\leq r_{i}$ for all $i=1,...,k$
the first condition of the monotonicity of the capacity transfers
implies that $q_{s}^{k+1}(r_{1},...,r_{k})\geq q_{s}^{k+1}(\tilde{r}_{1},...,\tilde{r}_{k})$.
On the other hand, the second condition of the monotonicity of the
capacity transfers implies that $q_{s}^{k+1}(r_{1},...,r_{k})\leq q_{s}^{k+1}(\tilde{r}_{1},...,\tilde{r}_{k})$. 

Since $x\notin Y_{k}$, we will have $x\notin\tilde{Y}_{k}$ for all
$k=1,...,\lambda_{s}$. Thus, we can conclude that $x\notin\overline{C}^{s}(Y\cup\{z\},f^{s},q_{s})$,
which tells us that the completion $\overline{C}^{s}$ satisfies the
substitutability condition.
\begin{prop}
$\overline{C}^{s}(\cdot)$ satisfies the LAD. 
\end{prop}

\paragraph{Proof. }

Consider two sets of contracts $Y$ and $\tilde{Y}$ such that $Y\subseteq\tilde{Y}\subseteq X$.
Let $f^{s}$ and $q_{s}$ be the precedence sequence and the capacity
transfer scheme of school $s\in S$. We want to show that 
\[
\mid\overline{C}^{s}(Y,f_{s},q_{s})\mid\leq\mid\overline{C}^{s}(\tilde{Y},f^{s},q_{s})\mid.
\]

Let $(Y_{j},r_{j},Y^{j})$ be the sequences of sets of chosen contracts,
numbers of vacant slots and sets of remaining contracts for groups
$j=1,...,\lambda_{s}$ under choice processes beginning with $Y=Y^{0}$.
Similarly, let $(\tilde{Y}_{j},\tilde{r}_{j},\tilde{Y}^{j})$ be the
sequences of sets of chosen contracts, numbers of vacant slots and
sets of remaining contracts for groups $j=1,...,\lambda_{s}$ under
choice processes beginning with $\tilde{Y}^{0}=\tilde{Y}$. 

For the first group with capacity $\overline{q}_{s}^{1}$, since the
sub-choice function is q-responsive (and thus implies the LAD), we
have 
\[
\mid Y_{1}\mid=\mid c_{1}^{s}(Y^{0},\overline{q}_{s}^{1},f^{s}(1))\mid\leq\mid c_{1}^{s}(\tilde{Y}^{0},\overline{q}_{s}^{1},f^{s}(1))\mid=\mid\tilde{Y}_{1}\mid.
\]
Then, it implies that $r_{1}=\overline{q}_{s}^{1}-\mid Y_{1}\mid\geq\tilde{r}_{1}=\overline{q}_{s}^{1}-\mid\tilde{Y}_{1}\mid$.
Moreover, we have $Y^{1}\subseteq\tilde{Y}^{1}$. To see this, consider
a $y\in Y^{1}$. It means that $y\notin Y_{1}$. If $y$ is not chosen
from a smaller set $Y^{0}$, then it cannot be chosen from a larger
set $\tilde{Y}^{0}$ because sub-choice function is q-responsive (hence,
substitutable). 

Suppose that $\tilde{r}_{j}\leq r_{j}$ and $Y^{j}\subseteq\tilde{Y}^{j}$
hold for all $j=1,...,k-1$. We need to show that both of them hold
for group $k$. 

Given that $\tilde{r}_{j}\leq r_{j}$ for all $j=1,...,k-1$, the
first condition of the monotonicity implies that $q_{s}^{k}(r_{1},...,r_{k-1})\geq q_{s}^{k}(\tilde{r}_{1},...,\tilde{r}_{k-1})$.
The second condition of the monotonicity puts an upper bound for the
difference between $q_{s}^{k}(r_{1},...,r_{k-1})$ and $q_{s}^{k}(\tilde{r}_{1},...,\tilde{r}_{k-1})$.
For group $k$ 
\[
\mid Y_{k}\mid-\mid\tilde{Y}_{k}\mid\leq\mid Y_{k}\mid-\mid c_{k}^{s}(Y^{k-1},q_{s}^{k}(\tilde{r}_{1},...,\tilde{r}_{k-1},f^{s}(k))\mid
\]
 because 
\[
\mid\tilde{Y}_{k}\mid=\mid c_{k}^{s}(\tilde{Y}^{k-1},q_{s}^{k}(\tilde{r}_{1},...,\tilde{r}_{k-1}),f^{s}(k))\mid\geq\mid c_{k}^{s}(Y^{k-1},q_{s}^{k}(\tilde{r}_{1},...,\tilde{r}_{k-1}),f^{s}(k))\mid
\]
 by the q-responsiveness of the sub-choice function. We then have
\[
\mid Y_{k}\mid-\mid c_{k}^{s}(Y^{k-1},q_{s}^{k}(\tilde{r}_{1},...,\tilde{r}_{k-1},f^{s}(k))\mid\leq q_{s}^{k}(r_{1},...,r_{k-1})-q_{s}^{k}(\tilde{r}_{1},...,\tilde{r}_{k-1}).
\]
 This follows from q-responsiveness because $\mid Y_{k}\mid-\mid c_{k}^{s}(Y^{k-1},q_{s}^{k}(\tilde{r}_{1},...,\tilde{r}_{k-1},f^{s}(k))\mid$
is the difference between the number of chosen contracts when the
capacity is (weakly) increased from $q_{s}^{k}(\tilde{r}_{1},...,\tilde{r}_{k-1})$
to $q_{s}^{k}(r_{1},...,r_{k-1})$. Hence, the difference $\mid Y_{k}\mid-\mid c_{k}^{s}(Y^{k-1},q_{s}^{k}(\tilde{r}_{1},...,\tilde{r}_{k-1},f^{s}(k))\mid$
cannot exceed the increase in the capacity which is $q_{s}^{k}(r_{1},...,r_{k-1})-q_{s}^{k}(\tilde{r}_{1},...,\tilde{r}_{k-1})$.
Therefore, now we have 
\[
\mid Y_{k}\mid-\mid\tilde{Y}_{k}\mid\leq q_{s}^{k}(r_{1},...,r_{k-1})-q_{s}^{k}(\tilde{r}_{1},...,\tilde{r}_{k-1}).
\]
 Rearranging gives us 
\[
q_{s}^{k}(\tilde{r}_{1},...,\tilde{r}_{k-1})-\mid\tilde{Y}_{k}\mid\leq q_{s}^{k}(r_{1},...,r_{k-1})-\mid Y_{k}\mid,
\]
 which is $\tilde{r}_{k}\leq r_{k}$. 

Given that $Y^{k-1}\subseteq\tilde{Y}^{k-1}$ and $q_{s}^{k}(r_{1},...,r_{k-1})\geq q_{s}^{k}(\tilde{r}_{1},...,\tilde{r}_{k-1})$,
we will have $Y^{k}\subseteq\tilde{Y}^{k}$. For an explanation, consider
a contract $x\in Y^{k}$. That means that $x\in Y^{k-1}$ but $x$
is not chosen from $Y^{k-1}$ when the capacity is $q_{s}^{k}(r_{1},...,r_{k-1})$,
i.e., $x\notin c_{k}^{s}(Y^{k-1},q_{s}^{k}(r_{1},...,r_{k-1}),f^{s}(x))$.
When the capacity is reduced to $q_{s}^{k}(\tilde{r}_{1},...,\tilde{r}_{k-1})$
and the set $Y^{k-1}$ is expanded to $\tilde{Y}^{k-1}$, $x$ cannot
be chosen because the sub-choice function is q-responsive. Hence,
it must be the case that $x\in\tilde{Y}^{k}$. 

Now let $\eta_{j}=r_{j}-\tilde{r}_{j}$. As we just proved above,
$\eta_{j}\geq0$ for all $j=1,...,\lambda_{s}$. Plugging $r_{j}=q_{s}^{j}(r_{1},...,r_{j-1})-\mid Y_{j}\mid$
and $\tilde{r}_{j}=q_{s}^{k}(\tilde{r}_{1},...,\tilde{r}_{k-1})-\mid\tilde{Y}_{j}\mid$
in $\eta_{j}=r_{j}-\tilde{r}_{j}$ gives us 
\[
\mid\tilde{Y}_{j}\mid=q_{s}^{j}(r_{1},...,r_{j-1})-q_{s}^{j}(\tilde{r}_{1},...,\tilde{r}_{j-1})+\mid Y^{j}\mid+\eta_{j}.
\]
Summing both the right and left hand sides for $j=1,...,\lambda_{s}$
yields 
\[
\stackrel[j=1]{\lambda_{s}}{\sum}\mid\tilde{Y}_{j}\mid=\stackrel[j=1]{\lambda_{s}}{\sum}\mid Y_{j}\mid+\stackrel[j=2]{\lambda_{s}}{\sum}\left[q_{s}^{j}(r_{1},...,r_{j-1})-q_{s}^{j}(\tilde{r}_{1},...,\tilde{r}_{j-1})\right]+\stackrel[j=1]{\lambda_{s}}{\sum}\eta_{j}.
\]
 Since each $\eta_{j}\geq0$, we have 
\[
\stackrel[j=1]{\lambda_{s}}{\sum}\mid\tilde{Y}_{j}\mid\geq\stackrel[j=1]{\lambda_{s}}{\sum}\mid Y_{j}\mid+\stackrel[j=2]{\lambda_{s}}{\sum}\left[q_{s}^{j}(r_{1},...,r_{j-1})-q_{s}^{j}(\tilde{r}_{1},...,\tilde{r}_{j-1})\right].
\]
 Also, we know that $q_{s}^{j}(r_{1},...,r_{j-1})\geq q_{s}^{j}(\tilde{r}_{1},...,\tilde{r}_{j-1})$
for all $j=2,...,\lambda_{s}$ by the first condition of the monotonicity
of the capacity transfer scheme as, $r_{i}\geq\tilde{r}_{i}$ for
all $i=1,...,j-1$ (Notice that for $j=1$, the capacity is fixed
to $\overline{q}_{s}^{1}$ under both processes.) Therefore, we have
\[
\stackrel[j=1]{\lambda_{s}}{\sum}\mid\tilde{Y}_{j}\mid\geq\stackrel[j=1]{\lambda_{s}}{\sum}\mid Y_{j}\mid,
\]
 which means $\mid\overline{C}^{s}(Y,f^{s},q_{s})\mid\leq\mid\overline{C}^{s}(\tilde{Y},f^{s},q_{s})\mid$.

\subsection{Proofs of Theorems}

\subsubsection*{Proof of Theorem 1}

In Proposition 1 we showed that each dynamic reserve choice function
has a completion. Propositions 2 and 3 show that the completion satisfies
the IRC and substitutability conditions, respectively. Then, by Theorem
2 of \emph{Hatfield and Kominers} (2019), there exists a stable outcome
with respect to the profile of schools' choice functions. 

\subsubsection*{Proof of Theorem 2 }

In Proposition 4 we showed that the substitutable completion satisfies
the LAD. Then, by the Theorem 3 of \emph{Hatfield and Kominers} (2019),
the COM is (weakly) group strategy-proof for students. 

\subsubsection*{Proof of Theorem 3}

Assume, toward a contradiction, that the COM does not respect unambiguous
improvements. Then, there exists a student $i\in I$, a preference
profile of students $P\in\times_{i\in I}\mathcal{P}^{i}$, and priority
profiles $\overline{\Pi}$ and $\Pi$ such that $\overline{\Pi}$
is an unambiguous improvement over $\Pi$ for student $i$ and 
\[
\varphi_{i}(P;\Pi)P^{i}\varphi_{i}(P;\overline{\Pi}).
\]
 Let $\varphi_{i}(P;\Pi)=x$ and $\varphi_{i}(P;\overline{\Pi})=\overline{x}$.
Consider a preference $\widetilde{P}^{i}$ of student $i$ according
to which the only acceptable contract is $x$, i.e., $\widetilde{P}^{i}:\:x-\emptyset_{i}$.
Let $\widetilde{P}=(\widetilde{P}^{i},P_{-i})$. We will first prove
the following claim:

\paragraph*{Claim: $\varphi_{i}(\widetilde{P};\Pi)=x$ $\protect\Longrightarrow$
$\varphi_{i}(\widetilde{P};\overline{\Pi})=x$. }

\paragraph*{Proof of the Claim: }

Consider the outcome of the COM under priority profile $\Pi$ given
the preference profile of students $\widetilde{P}$. Recall that the
order in which students make offers has no impact on the outcome of
the COP. We can thus completely ignore student $i$ and run the COP
until it stops. Let $Y$ be the resulting set of contracts. At this
point, student $i$ makes an offer for his only contract $x$. This
might create a chain of rejections, but it does not reach student
$i$. So, his contract $x$ is chosen by $\mathbf{s}(x)$ by, say,
the group $k$ with respect to the precedence sequence $f^{\mathbf{s}(x)}$
of school $\mathbf{s}(x)$. Now consider the COP  under priority profile
$\overline{\Pi}$. Again, we completely ignore student $i$ and run
the COP until it stops. The same outcome $Y$ is obtained, because
the only difference between the two COPs  is student $i$'s position
in the priority rankings. At this point, student $i$ makes an offer
for his only contract $x$. If $x$ is chosen by the same group $k$,
then the same rejection chain (if there was one in the COP  under
the priority profile $\Pi$) will occur and it does not reach student
$i$; otherwise, we would have a contradiction with the case under
priority profile $\Pi$. The only other possibility is the following:
since student $i$'s ranking is now (weakly) better under $\overline{\pi}^{\mathbf{s}(x)}$
compared to $\pi^{\mathbf{s}(x)}$, his contract $x$ might be chosen
by group $l<k$. Then, it must be the case that $r_{l}=0$ in the
COP  under both priority profiles $\Pi$ and $\overline{\Pi}$. Therefore,
by selecting $x$, the group $l$ must reject some other contract.
Let us call this contract $y$. If no contract of student $\mathbf{i}(y)=j$
is chosen between groups $l$ and $k$, then, by the q-responsiveness
of sub-choice functions, the groups' chosen sets between $l$ and
$k$ under both priority profiles are the same. Hence, the number
of remaining slots would be the same. In this case, $y$ is chosen
in the group $k$. Thus, if a rejection chain starts, it will not
reach student $i$; otherwise, we could have a contradiction due to
the fact that $x$ was chosen at the end of the COP  under priority
profile $\Pi$. A different contract of student $j$ cannot be chosen
between groups $l$ and $k$; otherwise, the observable substitutability\footnote{Dynamic reserves choice functions satisfy \emph{observable substitutability
}condition of \emph{Hatfield et al. } (2019). We refer readers to
\emph{Hatfield et al. } (2019) for the definitions of observable offer
processes and observable substitutability. Since dynamic reserves
choice functions have substitutable completion that satisfies the
size monotonicity, it satisfies observable substitutability. } of dynamic reserves choice function of school $\mathbf{s}(x)$ would
be violated. Therefore, if any contract of student $j$ is chosen
by these groups between $l$ and $k$, it must be $y$. If $y$ is
chosen by a group that precedes $k$, then it must replace a contract---we
call it $z$. By the same reasoning, no other contract of student
$\mathbf{i}(z)$ can be chosen before group $k$; otherwise, we would
violate the observable substitutability of the dynamic reserve choice
function of school $\mathbf{s}(x)$. Proceeding in this fashion leads
the same contract in group $k$ to be rejected and initiates the same
rejection chain that occurs under priority profile $\Pi$. Since the
same rejection chain does not reach student $i$ under priority profile
$\Pi$, it will not reach student $i$ under priority profile $\overline{\Pi}$,
which ends our proof for the claim. 

Since $\varphi_{i}(P;\Pi)=x$ and $\varphi_{i}(P;\overline{\Pi})=\overline{x}$
such that $xP^{i}\overline{x}$, if student $i$ misreports and submits
$\widetilde{P}^{i}$ under priority profile $\overline{\Pi}$ , then
she can successfully manipulate the COM. This is a contradiction because
we have already established that the COM is strategy-proof.

\subsubsection*{Proof of Theorem 4}

Consider school $s\in S$ with a precedence sequence $f^{s}$ and
a target capacity vector $(\overline{q}_{s}^{1},...,\overline{q}_{s}^{\lambda_{s}})$.
Let $\widetilde{q}_{s}$ and $q_{s}$ be two capacity transfer schemes
that are compatible with the precedence sequence $f^{s}$ and the
target capacity vector $(\overline{q}_{s}^{1},...,\overline{q}_{s}^{\lambda_{s}})$.
Suppose that the following two conditions hold:
\begin{itemize}
\item there exists $l\in\{2,...,\lambda_{s}\}$ and $(\hat{r}_{1},...,\hat{r}_{l-1})\in\mathbb{Z}_{+}^{l-1}$,
such that $\widetilde{q}_{s}^{l}(\hat{r}_{1},...,\hat{r}_{l-1})=1+q_{s}^{l}(\hat{r}_{1},...,\hat{r}_{l-1})$,
and
\item for all $j\in\{2,...,\lambda_{s}\}$ and $(r_{1},...,r_{j-1})\in\mathbb{Z}_{+}^{j-1}$,
if $j\neq l$ or $(r_{1},...,r_{j-1})\neq(\hat{r}_{1},...,\hat{r}_{l-1})$,
then $\widetilde{q}_{s}^{j}(r_{1},...,r_{j-1})=q_{s}^{j}(r_{1},...,r_{j-1})$. 
\end{itemize}
Let $\widetilde{C}^{s}$ and $C^{s}$ be dynamic reserves choice functions
$\widetilde{C}^{s}(\cdot,f^{s},\widetilde{q}_{s})$ and $C^{s}(\cdot,f^{s},q_{s})$,
respectively. Let $\widetilde{C}=\left(\widetilde{C}^{s},C_{-s}\right)$
and $C=\left(C^{s},C_{-s}\right)$. Let the outcomes of the cumulative
offer algorithm at $\left(P,\widetilde{C}\right)$ and $\left(P,C\right)$
be $\widetilde{Z}$ and $Z$, respectively. If $\widetilde{Z}=Z$,
then there is nothing to prove because it means the capacity flexibility
of school $s$ does not bite. 

Suppose that $\widetilde{Z}\neq Z$. That is, the capacity flexibility
of school $s$ bites, which means that there is a student who was
rejected under $C^{s}$ who is no longer rejected under $\widetilde{C}^{s}$.
We now define an \emph{improvement chains} algorithm that starts with
outcome $Z$. Since the capacity flexibility bites, the vector $(\hat{r}_{1},...,\hat{r}_{l-1})$
must occur in the choice procedure of school $s$. 

\paragraph*{Step 1: }

Consider students who prefer $(s,f^{s}(l))$ to their assignments
under $Z$, i.e., 
\[
\widetilde{I}_{1}^{(s,f^{s}(l))}=\{i\in I\mid(s,f^{s}(l))P^{i}Z_{i}\}.
\]

We choose $\pi^{s}$-maximal student in $\widetilde{I}_{1}^{(s,f^{s}(l))}$
(if any), call her $\widetilde{i}_{1}$, and assign her $\widetilde{x}_{1}=(\widetilde{i}_{1},s,f^{s}(l))$.
Update the outcome to $\widetilde{Z}_{1}=Z\cup\{\widetilde{x}_{1}\}\setminus z_{1}$
where $z_{1}$ is the contract student $\widetilde{i}_{1}$ receives
under $Z$. 

If $(\mathbf{s}(z_{1}),\mathbf{t}(z_{1}))=\emptyset$, then the improvement
process ends and we have $\widetilde{Z}=\widetilde{Z}_{1}=Z\cup\{\widetilde{x}_{1}\}$.
Otherwise, we move to Step 2 because by assigning $\widetilde{i}_{1}$
to $(s,f^{s}(l))$ we create a vacancy in school $\mathbf{s}(z_{1})$
within the privilege $\mathbf{t}(z_{1})$. 

If $\widetilde{I}_{1}^{(s,f^{s}(l))}=\emptyset$, then the number
of vacant slots at the\emph{ last} group accepting students in type
$f^{s}(l)$ will increase by one. When the capacity transfer scheme
of school $s$ does not transfer this extra vacancy to any other group
following the last group in type $f^{s}(l)$ in the computation of
$C^{s}(Z_{s},f^{s},\widetilde{q}_{s})$, the improvement chain process
ends and we have $\widetilde{Z}=Z$. If the extra slot is transferred
to the group $l^{'}$ that follows the last group in type $f^{s}(l)$
in the computation of $C^{s}(Z_{s},f^{s},\widetilde{q}_{s})$, then
we consider students who prefer $(s,f^{s}(l^{'}))$ over their assignments
under $Z$, i.e., 
\[
I_{1}^{(s,f^{s}(l^{'}))}=\{i\in I\mid(s,f^{s}(l^{'}))P^{i}Z_{i}\}.
\]
 We choose $\pi^{s}$-maximal student in $I_{1}^{(s,f^{s}(l^{'}))}$
(if there is any), call her $\widetilde{i}_{1}$, and assign her $\widetilde{x}_{1}=(\widetilde{i}_{1},s,f^{s}(l^{'}))$.
Update the outcome to $\widetilde{Z}_{1}=Z\cup\{\widetilde{x}_{1}\}\setminus z_{1}$
where $z_{1}$ is the contract $\widetilde{i}_{1}$ receives under
$Z$. 

If $(\mathbf{s}(z_{1}),\mathbf{t}(z_{1}))=\emptyset$, then the improvement
process ends and we have $\widetilde{Z}=\widetilde{Z}_{1}=Z\cup\{\widetilde{x}_{1}\}$.
Otherwise, we move to Step 2. Because assigning $\widetilde{i}_{1}$
to $(s,f^{s}(l^{'}))$ creates a vacancy in school $\mathbf{s}(z_{1})$
within the privilege $\mathbf{t}(z_{1})$. 

If $\widetilde{I}_{1}^{(s,f^{s}(l^{'}))}=\emptyset$, then the number
of vacant slots at the last group that accepts students in type $f^{s}(l^{'})$
will increase by one. If the capacity transfer scheme of school $s$
does not transfer this extra vacancy to any other group following
the last group that accepts students of type $f^{s}(l^{'})$ in the
computation of $C^{s}(Z_{s},f^{s},\widetilde{q}_{s})$, then the improvement
chain process ends and we have $\widetilde{Z}=Z$. If the extra slot
is transferred to the group $l^{''}$ that follows the last group
that accepts students in type $f^{s}(l^{'})$ in the computation of
$C^{s}(Z_{s},f^{s},\widetilde{q}_{s})$, then we consider students
who prefer $(s,f^{s}(l^{''}))$ over their assignments under $Z$,
and so on.

Since school $s$ has finitely many groups, Step 1 ends in finitely
many iterations. If no extra student is assigned to school $s$ by
the end of Step 1, then the improvement chains algorithm ends and
we have $\widetilde{Z}=Z$. If an extra student is assigned to school
$s$ by the end of Step 1, then we move on to Step 2. 

\paragraph*{Step t>1:}

Consider students who prefer $(\mathbf{s}(z_{t-1}),\mathbf{t}(z_{t-1}))$
to their assignments under $\widetilde{Z}_{t-1}$, i.e., 
\[
\widetilde{I}_{t}^{(\mathbf{s}(z_{t-1}),\mathbf{t}(z_{t-1}))}=\{i\in I\mid(\mathbf{s}(z_{t-1}),\mathbf{t}(z_{t-1}))P^{i}(\widetilde{Z}_{t-1})_{i}\}.
\]

We choose $\pi^{\mathbf{s}(z_{t-1})}$-maximal student in $\widetilde{I}_{t}^{(\mathbf{s}(z_{t-1}),\mathbf{t}(z_{t-1}))}$,
call her $\widetilde{i}_{t}$, and assign her $\widetilde{x}_{t}=(\widetilde{i}_{t},\mathbf{s}(z_{t-1}),\mathbf{t}(z_{t-1}))$.
Update the outcome to $\widetilde{Z}_{t}=\widetilde{Z}_{t-1}\cup\{\widetilde{x}_{t}\}\setminus z_{t}$
where $z_{t}$ is the contract student $\widetilde{i}_{t}$ receives
under $\widetilde{Z}_{t-1}$. 

If $(\mathbf{s}(z_{t-1}),\mathbf{t}(z_{t-1}))=\emptyset$, then the
improvement algorithm ends and we have $\widetilde{Z}=\widetilde{Z}_{t}=\widetilde{Z}_{t-1}\cup\{\widetilde{x}_{t}\}$.
Otherwise, we move to Step $t+1$. Because assigning $\widetilde{i}_{t}$
to $(\mathbf{s}(z_{t-1}),\mathbf{t}(z_{t-1}))$ creates a vacancy
in school $\mathbf{s}(z_{t}$) within type $\mathbf{t}(z_{t})$. 

If $\widetilde{I}_{t}^{(\mathbf{s}(z_{t-1}),\mathbf{t}(z_{t-1}))}=\emptyset$,
then the number of vacant slots at the last\emph{ }group that accepts
students in type $f^{\mathbf{s}(z_{t-1})}$ will increase by one.
If the capacity transfer scheme of school $\mathbf{s}(z_{t-1})$ does
not transfer this extra capacity to any other group following the
last group that accepts students in type $\mathbf{t}(z_{t-1})$ in
the computation of $C^{\mathbf{s}(z_{t-1})}((\widetilde{Z}_{t-1})_{\mathbf{s}(z_{t-1})},f^{\mathbf{s}(z_{t-1})},q_{\mathbf{s}(z_{t-1})})$,
then the improvement chains process ends and we have $\widetilde{Z}=\widetilde{Z}_{t-1}$.
If the extra slot is transferred to the group of slot $m$ that follows
the last group that accepts students in type $\mathbf{t}(z_{t-1})$
in the computation of $C^{\mathbf{s}(z_{t-1})}((\widetilde{Z}_{t-1})_{\mathbf{s}(z_{t-1})},f^{\mathbf{s}(z_{t-1})},q_{\mathbf{s}(z_{t-1})})$,
then we consider students who prefer $(\mathbf{s}(z_{t-1}),f^{\mathbf{s}(z_{t-1})}(m))$
over their assignments under $\widetilde{Z}_{t-1}$, i.e., 
\[
\widetilde{I}_{t}^{(\mathbf{s}(z_{t-1}),f^{\mathbf{s(z_{t-1})}}(m))}=\{i\in I\mid(\mathbf{s}(z_{t-1}),f^{\mathbf{s}(z_{t-1})}(m))P^{i}(\widetilde{Z}_{t-1})_{i}\}.
\]
 We choose $\pi^{\mathbf{s}(z_{t-1})}$-maximal student in $\widetilde{I}_{t}^{(\mathbf{s}(z_{t-1}),f^{\mathbf{s}(z_{t-1})}(m))}$,
call her $\widetilde{i}_{t}$, and assign her $\widetilde{x}_{t}=(\widetilde{i}_{t},\mathbf{s}(z_{t-1}),f^{\mathbf{s}(z_{t-1})}(m))$.
Update the outcome to $\widetilde{Z}_{t}=\widetilde{Z}_{t-1}\cup\{\widetilde{x}_{t}\}\setminus z_{t}$
where $z_{t}$ is the contract student $\widetilde{i}_{t}$ receives
under $\widetilde{Z}_{t-1}$. 

If $(\mathbf{s}(z_{t-1}),f^{\mathbf{s}(z_{t-1})}(m))=\emptyset$,
then the improvement algorithm ends and we have $\widetilde{Z}=\widetilde{Z}_{t}=\widetilde{Z}_{t-1}\cup\{\widetilde{x}_{t}\}$.
Otherwise, we move to Step $t+1$. Because assigning $\widetilde{i}_{t}$
to $(\mathbf{s}(z_{t-1}),f^{\mathbf{s}(z_{t-1})}(m))$ creates a vacancy
in school $\mathbf{s}(z_{t}$) within type $\mathbf{t}(z_{t})$. 

If $\widetilde{I}_{t}^{(\mathbf{s}(z_{t-1}),\mathbf{t}(z_{t-1}))}=\emptyset$,
then the number of vacant slots at the last\emph{ }group that accepts
students in type $f^{\mathbf{s}(z_{t-1})}$ will increase by one.
If the capacity transfer scheme of school $\mathbf{s}(z_{t-1})$ does
not transfer this extra capacity to any other group following the
last group that accepts students in type $f^{\mathbf{s}(z_{t-1})}(m)$
in the computation of $C^{\mathbf{s}(z_{t-1})}((\widetilde{Z}_{t-1})_{\mathbf{s}(z_{t-1})},f^{\mathbf{s}(z_{t-1})},q_{\mathbf{s}(z_{t-1})})$,
then the improvement chains process ends and we have $\widetilde{Z}=\widetilde{Z}_{t-1}$.
If the extra slot is transferred to the group of slot $m^{'}$ that
follows the last group that accepts students in type $f^{\mathbf{s}(z_{t-1})}(m)$
in the computation of $C^{\mathbf{s}(z_{t-1})}((\widetilde{Z}_{t-1})_{\mathbf{s}(z_{t-1})},f^{\mathbf{s}(z_{t-1})},q_{\mathbf{s}(z_{t-1})})$,
then we consider students who prefer $(\mathbf{s}(z_{t-1}),f^{\mathbf{s}(z_{t-1})}(m^{'}))$
over their assignments under $\widetilde{Z}_{t-1}$, and so on. 

Since school $\mathbf{s}(z_{t-1})$ has finitely many groups , Step
$t$ ends in finitely many iterations. If no extra student is assigned
to school $\mathbf{s}(z_{t-1})$ by the end of Step $t$, then the
improvement chains algorithm ends and we have $\widetilde{Z}=\widetilde{Z}_{t-1}$.
If an extra student is assigned to school $\mathbf{s}(z_{t-1})$ by
the end of Step $t$, then we move on to Step $t+1$. 

This process ends in finitely many iterations because there are finitely
many contracts and when we move to the next step it means a student
is made strictly better off. Also, notice that no student is worse
off during the execution of the improvement chains algorithm. The
improvement algorithm, by construction, starts with the outcome $\Phi(P,C)$
and ends at $\Phi(P,\widetilde{C})$. Hence, we have $\Phi_{i}(P,\widetilde{C})R^{i}\Phi_{i}(P,C)$
for all $i\in I$. 

We define the sequence of capacity transfer schemes and dynamic reserve
choice functions for school $s\in S$: $\left((q_{s})^{1},(q_{s})^{2},...\right)$
and $\left(C^{s}(Y,f^{s},(q_{s})^{1}),C^{s}(Y,f^{s},(q_{s})^{2}),...\right)$.
Let the sequence $\Phi(P,C^{1})$, $\Phi(P,C^{2})$,... denote the
outcomes of the COPs  at profiles $(P,(C^{s}(\cdot,f^{s},(q_{s})^{1}),C_{-s}))$
and $(P,(C^{s}(\cdot,f^{s},(q_{s})^{2}),C_{-s}))$,..., respectively.
Hence, by construction, we have $\Phi_{i}(P,C^{a+1})R^{i}\Phi_{i}(P,C^{a})$
for all $i\in I$ and $a\geq1$. By the transitivity of weak preferences,
we have $\Phi_{i}(P,\widetilde{C})R^{i}\Phi(P,C)$ for all $i\in I$. 

\subsubsection*{Proof of Theorem 5}

Our proof is constructive. We first define an associated type space.
Let $X$ be the set of all contracts. We define a distinct ``type''
for each contract in $X$. Let $g:\:X\rightarrow\mathcal{T}=\{\tau_{1},...,\tau_{\mid X\mid}\}$
be a bijective function. The interpretation of the $g$ function is
that the artificial type of a contract $x\in X$ is $g(x)\in\{\tau_{1},...,\tau_{\mid X\mid}\}$.
Therefore, each contract in $X$ is associated with a distinct (artificial)
type. 

Consider a slot $b_{l}\in\mathcal{B}_{s}$ with priority order $\pi^{b_{l}}$.
Let $\mid\pi^{b_{l}}\mid$ denote the number of contracts that the
slot $b_{l}$ finds acceptable, i.e., ranks higher than the null contract
which corresponds to remaining unassigned. Let $x_{l}^{1}$, $x_{l}^{2}$,...,$x_{l}^{\mid\pi^{b_{l}}\mid}$
be the acceptable contracts for slot $b_{l}$ such that 
\[
x_{l}^{1}\pi^{b_{l}}x_{l}^{2}\pi^{b_{l}}\cdots\pi^{b_{l}}x_{l}^{\mid\pi^{b_{l}}\mid}.
\]
 For the slot $b_{l}$ in school $s$ in the true market, we create
a \emph{sequence of slots}---$\mid\pi^{b_{l}}\mid$ many slots---
in the associated market, i.e., $\{b_{l}^{1},...b_{l}^{\mid\pi^{b_{l}}\mid}\}$.
The initial capacity of $b_{l}^{1}$ is $1$, i.e., $\overline{q}_{b_{l}^{1}}=1$,
and the initial capacities of $b_{l}^{2},b_{l}^{3},...,b_{l}^{\mid\pi^{b_{l}}\mid}$
are $0$, i.e., $\overline{q}_{b_{l}^{k}}=0$ for all $k=2,...,\mid\pi^{b_{l}}\mid$.
Define $r_{b_{l}^{k}}$ such that $r_{b_{l}^{k}}=0$ if slot $b_{l}^{k}$
is filled and $r_{b_{l}^{k}}=1$ if slot $b_{l}^{k}$ remains vacant.
The dynamic capacity of the slot $b_{l}^{k}$, for all $k=2,...,\mid\pi^{b_{l}}\mid$,
is defined as $q_{b_{l}^{k}}(r_{b_{l}^{1}},...,r_{b_{l}^{k-1}})=r_{b_{l}^{k-1}}.$
That is, if the slot $b_{l}^{k-1}$ remains vacant, then the capacity
of the slot $b_{l}^{k}$ becomes 1. Note that if a slot $b_{l}^{k-1}$
is filled, then the dynamic capacity of slots that come after $b_{l}^{k-1}$
become $0$. 

Each slot $b_{l}^{k}$ is associated with a sub-choice rule $c_{b_{l}^{k}}^{s}(\cdot,q_{b_{l}^{k}},\cdot)$
that is defined as follows: The sub-choice rule $c_{b_{l}^{k}}^{s}(\cdot,q_{b_{l}^{k}},\cdot)$
can only considers contracts with artificial type $g^{-1}(x_{l}^{k})$,
therefore only the contract $x_{l}^{k}$. Given a set of contracts
$Y\subseteq X$, 
\[
c_{b_{l}^{k}}^{s}(Y,q_{b_{l}^{k}},g^{-1}(x_{l}^{k}))=\begin{cases}
\begin{array}{c}
\{x_{l}^{k}\}\\
\emptyset
\end{array} & \begin{array}{c}
if\;x_{l}^{k}\in Y\;and\;q_{b_{l}^{k}}=1,\\
otherwise
\end{array}.\end{cases}
\]
Note that $c_{b_{l}^{k}}^{s}$ is a q-responsive choice function.
We now describe a dynamic reserves choice rule $\widetilde{C}^{s}(\cdot)$
that is outcome equivalent to the slot-specific choice rule $C^{s}(\cdot)$.
Let $Y\subseteq X$ be a set of contracts. 

\paragraph*{Step 1}

Consider slots $\{b_{1}^{1},b_{1}^{2},...,b_{1}^{\mid\pi^{b_{1}}\mid}\}$
in this step. 

\subparagraph*{Step 1.1 }

Apply the sub-choice function $c_{b_{1}^{1}}^{s}$. If a contract
is chosen, then end Step 1, and move to Step 2 due to the capacity
transfer rule described above. Otherwise, move to Step 1.2.

\subparagraph*{Step 1.2 }

Apply the sub-choice function $c_{b_{1}^{2}}^{s}$. If a contract
is chosen, then end Step 1, and move to Step 2 due to the capacity
transfer rule described above. Otherwise, move to Step 1.3.

This process continues in sequence. If a contract chosen in Step 1,
then all of the contracts associated with the student whose contract
is chosen is removed for the rest of the procedure. Let $y^{1}$ be
the chosen contract in this step. Then, the set of remaining contracts
is $Y\setminus Y_{\mathbf{i}(y^{1})}$. 

\paragraph*{Step $\mathbf{n}\protect\geq2$ }

Consider slots $\{b_{n}^{1},b_{n}^{2},...,b_{n}^{\mid\pi^{b_{n}}\mid}\}$
in this step. 

\subparagraph*{Step n.1 }

Apply the sub-choice function $c_{b_{n}^{1}}^{s}$. If a contract
is chosen, then end Step n, and move to Step $(n+1)$ due to the capacity
transfer rule described above. Otherwise, move to Step n.2.

\subparagraph*{Step n.2 }

Apply the sub-choice function $c_{b_{n}^{2}}^{s}$. If a contract
is chosen, then end Step n, and move to Step $(n+1)$ due to the capacity
transfer rule described above. Otherwise, move to Step n.3.

This process continues in sequence. If a contract chosen in Step n,
then all of the contracts associated with the student whose contract
is chosen is removed for the rest of the procedure. Let $y^{n}$ be
the chosen contract in this step. Then, the set of remaining contracts
is $Y\setminus Y_{\mathbf{i}(y^{1},...,y^{n})}$. 

By construction, for any given set of contracts $Y\subseteq X$, for
each slot $b_{l}$ in the process of the slot-specific priorities
choice function $C^{s}(\cdot)$ and Step $l$ of the dynamic reserves
choice function $\widetilde{C}^{s}(\cdot)$ the set of available contracts,
and hence, the chosen contract are the same. Therefore, these two
choice functions select the same set of contracts, i.e., $C^{s}(Y)=\widetilde{C}^{s}(Y)$.
This ends our proof. 

\pagebreak
\end{document}